\documentclass[12pt]{article}
\usepackage[utf8]{inputenc} 
\usepackage[T1]{fontenc}    
\usepackage{hyperref}       
\usepackage{url}            
\usepackage{booktabs}       
\usepackage{amsfonts}       
\usepackage{nicefrac}       
\usepackage{microtype}      
\usepackage{xcolor}         
\usepackage{amsmath}
\usepackage{float}
\RequirePackage{graphicx}
\usepackage{graphicx}
\usepackage{caption}
\usepackage{subcaption}
\usepackage{relsize}
\usepackage{mathtools}
\usepackage{multirow}
\usepackage[toc,page]{appendix}
\usepackage{xcolor}
\usepackage{colortbl} 
\newcommand{\diff}{\mathrm{d}}
\newcommand\norm[1]{\lVert#1\rVert}
\newcommand{\difff}{\mathrm{~d}}

\newcommand{\blind}{0}

\newcommand{\review}[1]{#1}

\begin{document}



\def\spacingset#1{\renewcommand{\baselinestretch}%
{#1}\small\normalsize} \spacingset{1}


\if0\blind
{
  \title{\bf Estimating velocities of infectious disease spread through spatio-temporal log-Gaussian Cox point processes}
  \author{\textbf{Fernando Rodriguez Avellaneda} \\
Computer, Electrical and Mathematical Sciences and Engineering Division,\\
King Abdullah University of Science and Technology (KAUST),\\
Thuwal, Saudi Arabia,\\
    \textbf{Jorge Mateu} \\
    Department of Mathematics,\\
    Universitat Jaume I, Castellon, Spain \\
    and \\
    \textbf{Paula Moraga} \\
Computer, Electrical and Mathematical Sciences and Engineering Division,\\
King Abdullah University of Science and Technology (KAUST),\\
Thuwal, Saudi Arabia}
  \maketitle
} \fi

\if1\blind
{
  \bigskip
  \bigskip
  \bigskip
  \begin{center}
    {\LARGE\bf Estimating velocities of infectious disease spread through spatio-temporal log-Gaussian Cox point processes}
\end{center}
  \medskip
} \fi

\bigskip
\begin{abstract}
Understanding the spread of infectious diseases such as COVID-19 is crucial for informed decision-making and resource allocation. A critical component of disease behavior is the velocity with which disease spreads, defined as the rate of change between time and space. This paper proposes a spatio-temporal modeling approach to determine the velocities of infectious disease spread. Our approach assumes that the locations and times of people infected can be considered a spatio-temporal point pattern that arises as a realization of a spatio-temporal log-Gaussian Cox point process. The intensity function of this process is estimated using a fully non-separable spatio-temporal model derived from diffusion Stochastic Partial Differential Equations (SPDE), and fast Bayesian inference is performed using integrated nested Laplace approximation (INLA). The velocity is then calculated using finite differences that approximate the derivatives of the intensity function. Finally, the directions and magnitudes of the velocities can be mapped at specific times to better examine the spread of the disease throughout the region. This method is demonstrated by analyzing COVID-19 spread in Cali, Colombia, during the 2020-2021 pandemic.
\end{abstract}

\noindent%
{\it Keywords:}  Bayesian inference, Infectious diseases, INLA, Spatio-temporal modeling, Stochastic Partial Differential Equations

\spacingset{1.45} 
\section{Introduction}

Transmission of infectious diseases such as COVID-19 has the potential to collapse the healthcare system, causing health, societal, and economic problems \cite{kaye_et_al_2021} \cite{Ttianchen_et_al_2021}. For that reason, modeling these diseases is critical to understanding how they spread, and help decision-makers create policies to mitigate them. In this regard, a powerful tool in the context of spatio-temporal point patterns that can help understand the spread of an infectious disease is the \textit{velocity} that could be computed given a direction in any location in space and time. This magnitude could be understood as the instantaneous rate of change for the point pattern at location and time $(\textbf{u},t)$ in direction $\textbf{v}$.

In the context of point patterns, velocity is a quantitative measure that helps us to understand how a spatio-temporal process evolves. Moreover, when it is possible to define the intensity function, velocity refers to how fast the intensity changes in time relative to its change in a given spatial direction. Thus, this measure captures the speed of the intensity change in that direction, and provides information about the spatio-temporal dynamics of intensity and, consequently, the underlying point pattern. In this context, velocity could be interpreted in an \textit{instantaneous} sense as the chance for an event to lie in a small product neighborhood of a specific location and time. 

Our particular interest concerns the velocity of the spread of contagious diseases over space and time. That means we used the location of infected people by an infectious disease in a particular time resolution, such as days, weeks, or months, and a space window, such as a city or country. With that information, we can define a spatio-temporal point pattern $\{ \zeta_i \}_{i=1}^{n} = \{ \left(\textbf{u}_i,t_i \right) \}_{i=1}^{n}$, with the velocity in a given location $\textbf{u}_0$ and time $t_0$ in the direction $\textbf{v}$ by the instantaneous relative change for an infected individual in time over the instantaneous change in space in the direction $\textbf{v}$. Now, the direction of minimum velocity is associated with the direction of the largest instantaneous change for an event in space. This direction also provides the slowest speed of change in the chance of an event, as explained in Section \ref{sec:velocity}. 

Many authors implemented different models to estimate the directional derivatives of the intensity of a point pattern. For instance, \cite{Banerjee_2003} introduced a formal directional derivative process based on the realization of a stochastic process. Later, \cite{terres_gelfan_2016} developed a methodology for local sensitivity analysis based on directional derivatives that also uses continuous covariates to relate them with spatial gradients. For geostatistical data, temporal and spatial derivatives were computed to estimate velocities of climate change \cite{Schliep_et_al_2015} relating covariates to this, for example, temperature. Extending these ideas to point patterns \cite{schliep_gelfand_2019} compute velocities for spatio-temporal point patterns, assuming that the point pattern follows a log-Gaussian Cox (LGCP) process and also by using Nearest Neighbor Gaussian process (NNGP). Additionally, other authors consider other ways to compute velocities that use higher-order derivatives, such as second-order derivatives \cite{quick_et_al_2015}. 

\review{In the context of infectious disease spreading, particularly in COVID-19 in the region of Cali, Colombia, several approaches have been used to understand, model, and predict the spread around the city. For instance, \cite{ribeiroamaraletal23} estimates the number of infected individuals using a SIR model to predict future intensity patterns. However, this method does not provide an insightful estimation of the directions and magnitude of the disease propagation. On the other hand, \cite{dongetal2023} uses kernel-induced feature functions to understand how fast the disease spreads around the city. However, it does not provide direction or estimation of this change. In this paper, we propose a computation of velocities for spatio-temporal point patterns using finite differences as a way to understand the magnitude and direction of disease propagation in a specific region, as it described in Section \ref{sec:velocity-exa}.}

A key strength of our method is its flexibility: it does not assume a specific parametric form for the intensity function or a fixed covariance structure.  This allows us to detect instantaneous changes that parametric models may be too rigid to capture. Nevertheless, because our nonparametric approach is data-driven, it can be sensitive to noise, bias, or irregularities in the observed data. To mitigate this, we implement the fully nonseparable covariance structure described in \cite{lindgren2023inlaspacetime}, which helps reduce data dependence and improves the consistency of velocity estimates, mainly when the number of spatio-temporal points is low, as we explain in Section \ref{sec:sim}.

For inference, we employ the integrated nested Laplace approximation (INLA) framework \cite{rue_et_al_r_inla_2009} and \texttt{inlabru} \cite{inlabru}, which offers computational efficiency and simplicity for estimating the posterior distribution of the intensity function. Specifically, we use \texttt{INLAspacetime} \cite{lindgren2023inlaspacetime}, which extends the separable spatio-temporal covariance structure to a fully nonseparable through diffusion-based models.

The structure of the remainder of the paper is as follows. In Section \ref{sec:methods}, we introduce some theory about point processes and the log-Gaussian Cox process. Then, in Section \ref{sec:methodology}, we fully detail how we decompose the intensity function into three components and how inference for the LGCP is performed  through its covariance structure. In this section, we also explain how velocity is defined for spatio-temporal point patterns, and how to approximate it using finite differences. In Section \ref{sec:sim}, we present a simulation study to assess the performance of our method using a point pattern with a deterministic intensity function. Then, in Section \ref{sec:app}, we present an application using data from COVID-19 in Cali, Colombia. Finally, we present the conclusions, a discussion of the findings, and some directions for future research. 

\section{Preliminary concepts}
\label{sec:methods}

In this section, we first present some theoretical concepts related to point processes necessary to introduce the methodology used to model the spread of disease and to compute velocities. Using these concepts, we can provide a detailed explanation of how disease spread can be modeled both spatially and temporally. This involves analyzing the locations of infected individuals as they change over time. By tracking these observed locations over a specific period, we can gain valuable insights into the patterns and dynamics of disease transmission. This approach allows us to understand where infections occur and how they evolve, which is crucial for effective disease control and prevention.

\subsection{Point processes}

Given a subset $W \subseteq \mathbb{R}^2$ and an interval $T \subseteq \mathbb{R}$, a spatio-temporal point pattern consists of a collection of all points $\{ \zeta_i \}_{i=1}^{n} = \{ \left(\textbf{u}_i,t_i \right) \}_{i=1}^{n} \subseteq W \times T$. Under these conditions, a spatio-temporal point process is a random mechanism whose outcome or realization is a point pattern. For a spatio-temporal point process we define the intensity function $\lambda: W \times T \to [0,\infty)$ such that $\int_{A} \int_{B} \lambda(\textbf{u},t) \diff t \difff \textbf{u} < \infty$ for all bounded sets $A \subseteq W$, and intervals $B \subseteq T$ that satisfies the following property
\begin{equation*}
\mathbb{E}\left[\mathcal{N}\left(A,B\right)\right]=\int_{A} \int_{B} \lambda\left( \textbf{u},t\right) \diff t \difff \textbf{u}, 
\end{equation*}
where $\mathcal{N}\left(A,B\right)$ is the number of points in the region $A$ in the time interval $B$. Note that we assume that $T$ is an interval in $\mathbb{R}$, but we will often have $T \subseteq \mathbb{Z}$. In this scenario, we can consider the event times $t_i$ as marks of a spatial point pattern with location $\textbf{u}_i$. 

One of the most important point process models is the Poisson process. This process is frequently used as a benchmark due to its simplicity, and satisfies the following conditions:

\begin{enumerate}
    \item For any bounded sets $A \subseteq \mathbb{R}^2$, and interval $B \subseteq \mathbb{R}$, the number of points in the region $A$ in the interval $B$, $\mathcal{N}\left(A,B\right)$ follows a Poisson distribution with mean $\int_{A} \int_{B} \lambda\left(\textbf{u},t\right) \diff t \difff \textbf{u}$.
    \item For any bounded set $A$, any interval $B$ and any $n \in \mathbb{N}$, conditional on $\mathcal{N}\left(A,B\right)=n$, the $n$ events in $W \times T$ constitute an independent random sample drawn from the distribution over $W \times T$, which is characterized by the density function
    
    $$f \left( \textbf{u},t \right) = \frac{\lambda(\textbf{u},t)}{\int_{W} \int_{T} \lambda\left(\textbf{u},t\right) \diff t \difff \textbf{u}}.$$
\end{enumerate}

A Poisson process is described by its intensity function $\lambda(\textbf{u},t)$, and when this function is constant over the space-time domain, the process is homogeneous. Otherwise, the process is inhomogeneous. 

\subsection{Log-Gaussian Cox processes}

Log-Gaussian Cox processes (LGCPs) are typically used to model complex relationships between space and time, such as the spread of a contiguous disease \cite{diggle_et_al_2013}. To understand this complex process, we start by discussing the Cox process. A Cox process can be viewed as a Poisson process with a random intensity function. That is, the intensity function $\Lambda(\textbf{u},t)$ meets these two postulates:

\begin{enumerate}
    \item $\Lambda(\textbf{u},t)$ with  $\textbf{u} \in W$ and $t \in T$ is a non-negative valued stochastic process.
    \item Conditional on the realization $\Lambda(\textbf{u},t) = \lambda(\textbf{u},t)$ with $\textbf{u} \in W$ and $t \in T$, the point process is an inhomogeneous Poisson process with intensity $\lambda(\textbf{u},t)$.  
\end{enumerate}

LGCP models were introduced as a particular case of a Cox process \cite{mollet_et_al_1998}. As the name suggests, an LGCP is a Cox process with intensity function given by a Gaussian process, that is $\log (\Lambda(\mathbf{u},t)) = \mu(\textbf{u},t) + \xi(\mathbf{u},t)$, where $\exp(\mu(\textbf{u},t))$ could be interpreted as the mean structure of $\Lambda(\textbf{u},t)$ and $\xi(\mathbf{u},t)$ is a stationary Gaussian process.

\section{Methodology}
\label{sec:methodology}

To estimate velocities of an infectious disease, we propose a spatio-temporal point process model to describe how infectious diseases spread in space across time. Specifically, we use an LGCP model where the intensity is multiplicatively decomposed into temporal, spatial, and spatio-temporal stochastic variation as presented in \cite{diggle_et_al_2005} and \cite{ribeiroamaraletal23}. Once the model is fitted, the intensity estimate is used to compute \textit{minimal velocities} through finite differences. This approach allows us to understand the direction and velocity of disease spread.

\subsection{Spatio-temporal LGCP model}
\label{sec:lgpc_general}

The intensity function of a spatio-temporal LGCP model can be decomposed into three components, namely, a temporal component $\mu(t)$, a spatial component $\eta(\textbf{u})$, and a spatio-temporal stochastic variation $\xi(\textbf{u},t)$. This decomposition is represented as follows

\begin{equation}
    \label{ecu:geninten}
    \Lambda(\textbf{u},t)= \eta(\textbf{u}) \cdot \mu(t) \cdot \exp \left( \xi(\textbf{u},t) \right).
\end{equation}

Here, the spatial and temporal components are deterministic and capture aspects of the underlying population at risk for the disease pattern.
The spatial component $\eta(\textbf{u})$ represents the population density in the study region. 
This implies that if there were zones or locations with higher or lower cases, this variation would be absorbed in $\eta(\textbf{u})$.
The temporal component $\mu(t)$ is estimated by grouping the total number of cases over time and using a spline model with different basis functions to model these data. 
Thus, the flexibility of the splines estimation of $\mu(t)$ captures the evolution in the number of cases
over time. This emphasizes that the remaining term $\xi(\textbf{u},t)$ will capture anomalies in space and time.
This term is modeled as a stationary, zero-mean log-Gaussian stochastic process.
In the following sections, we explain in detail how we compute each component and their importance in the model.

\subsubsection{Temporal component}
\label{sec:temp}

The first step in our approach is to model the number of individuals per temporal unit (e.g., days, hours, etc.) infected by the disease. Accurately capturing these data is crucial for understanding the spread and making reliable predictions. We employed a spline model, a flexible statistical tool capable of fitting complex data patterns to achieve this goal.

Splines allow to handle non-linear trends in the data, providing a more precise fit than traditional linear models \cite{wood_2017}. This flexibility is essential in epidemiological studies, where the number of infections can vary greatly due to numerous factors such as periodic changes due to day-of-week effects or time-of-year events. We therefore estimate the number of infected people in time $t$, denoted by $\mu(t)$, by fitting the following standard log-spline model

\begin{equation*}
    \log(\mu(t)) = \sum_{i=1}^{q} \beta_i b_i(t),
\end{equation*}
where $b_i$ are completely known functions, commonly referred to as basis functions, and $q$ denotes the number of functions. Some examples of these basic functions are polynomials, roots, cosine, sine, and step functions. However, the election of the basic functions and their number highly depends on the data.

\subsubsection{Spatial component}

The main objective of this component is to approximate population density. This component is crucial for understanding the patterns of spread of an infectious disease because it directly influences how infections propagate through different regions. High-population-density areas may experience faster and more intense outbreaks due to the increased likelihood of contact between individuals.

Different methods could be used to compute the population density $\eta(\textbf{u})$ over a study region. In this work, we use population information from WorldPop \cite{world_pop_2020}, an open-access project that combines official census information and satellite images to estimate population density at different resolutions. For regions where population information is not available, we could estimate the population density using nonparametric methods such as Gaussian kernel estimators \cite{gonzalezandmoraga23}. For instance, the population density could be estimated as
\begin{equation}
    \label{ecu:kernel}
    \widehat{\eta}(\textbf{u})= \left( \frac{1}{n} \right) \mathlarger{\mathlarger{\sum}}_{i=1}^n \left( \frac{1}{h^2} \right) \phi\left(\frac{\textbf{u}-\textbf{u}_i} {h}\right),
\end{equation}
where $h>0$ is a fixed bandwidth, $\phi(\textbf{u}) = (2 \pi)^{-1} \exp{\{-0.5 \norm{\textbf{u}}^{2}\}}$, and $\textbf{u}_i$ are the locations of the infected individuals in the study region across time. We could also use an adaptive bandwidth $h_i$ for each location $\textbf{u}_i$ to have a better approximation.

\subsubsection{Spatio-temporal component}
\label{sec:spatemp}

Once the spatial and temporal components are estimated, they can be used to compute the intensity function of the spatio-temporal process. In this case, we assume that a log-Gaussian Cox process generates the observed point process. Then, the counting number of infected individuals in the region $W$ at the interval time $T$, denoted  $\mathcal{N}\left(W,T\right)$, meets the following condition:

\begin{equation*}
\mathcal{N}\left(W,T\right) | \Lambda(\textbf{u},t) = \lambda(\textbf{u},t)  \sim \text{Poisson} \left( \int_{W} \int_{T} \lambda\left( \textbf{u},t\right) \diff t \difff \textbf{u} \right),
\end{equation*}
where $\lambda\left( \textbf{u},t\right)$ can be decomposed as in (\ref{ecu:geninten}) as

\begin{equation}
\label{ecu:genintenexp}
    \Lambda(\textbf{u},t)= \eta(\textbf{u}) \cdot \mu(t) \cdot \exp \left( \xi(\textbf{u},t) \right).
\end{equation}

Here, the last term can be conveniently reparametrized as $\xi(\textbf{u},t) = \beta + \vartheta(\textbf{u},t)$ with $\vartheta(\textbf{u},t)$ following a zero-mean Gaussian process governed by a spatio-temporal diffusion Stochastic Partial Differential Equation (SPDE), which we introduce below.
Following the approach of \cite{lindgren2023inlaspacetime}, consider the operator $L_{u} = \gamma_u^2 - \Delta$, with $\gamma_u>0$ on a purely spatial domain $\mathcal{D} \subseteq \mathbb{R}^d$ on appropriated boundary conditions for compact domains. The generalized Whittle-Matérn covariance can be characterized through the precision operator $Q(\gamma_u,\gamma_e,\alpha) = \gamma_e^2 L_u^{\alpha}$, where the solution $\varphi(\textbf{u},t)$ satisfies the spatial SPDE

\begin{equation}
    \gamma_e L_u^{\alpha/2} \varphi(\textbf{u}) = \mathcal{W}(\textbf{u}) \quad \textbf{u} \in \mathcal{D},
\end{equation}
where $\mathcal{W}$ is a spatial white noise process. From this equation, we can define a noise process $\text{d} \mathcal{E}_{Q}(\textbf{u},t)$ as a Gaussian noise that is uncorrelated in time but correlated in space, with precision operator $Q=Q(\gamma_u,\gamma_e,\alpha_e)$ for a non-negative $\alpha_e$. With this noise process, these three non-negative smoothness parameters $( \alpha_t,\alpha_u,\alpha_e )$ and three positive scale parameters $(\gamma_t,\gamma_u,\gamma_e)$, we can define the following diffusion spatio-temporal SPDE

\begin{equation}
    \left(\gamma_t \frac{d}{d t}+L_u^{\alpha_u}\right)^{\alpha_t} v(\boldsymbol{u}, t)=\mathrm{d} \mathcal{E}_Q(\boldsymbol{u}, t), \quad(\boldsymbol{u}, t) \in \mathcal{D} \times \mathbb{R}.
    \label{ecu:diffusion}
\end{equation}

The solutions $\varphi(\textbf{u},t)$ of 
(\ref{ecu:diffusion}) define Gaussian processes with specific structural properties. Parameter $\alpha_e$ governs the degree of separability in the model, where setting $\alpha_e = 0$ leads to a fully nonseparable model. To further quantify this, \cite{lindgren2023inlaspacetime} introduce the nonseparability parameter  

\begin{equation*}
    \beta_u = 1 - \frac{\alpha_e}{\alpha} = 1 - \frac{\alpha_e}{v_u + d/2},
\end{equation*}
where $v_u$ represents the smoothness of the spatial process, and $\beta_u \in [0,1]$. The cases $\beta_u = 0$ and $\beta_u = 1$ correspond to fully separable and fully nonseparable models, respectively. Notably, as a special case, when $\alpha_u = 0$, it follows that $v_u = \alpha_e - d/2$, which directly implies $\beta_u = 0$, leading to a separable model. 

Then the solution of (\ref{ecu:diffusion}) for some smoothing parameters $(\alpha_t,\alpha_u,\alpha_e)$ is the residual normalized spatio-temporal component $\vartheta(\textbf{u},t)$ in (\ref{ecu:genintenexp}) that captures the underlying stochastic variation in the model.

\subsubsection{Inference}
\label{sec:inference}

In summary, the intensity function that models the infected individuals at time $t$ in the spatial window is defined as follows

\begin{align}
\begin{split}
    \Lambda(\textbf{u},t) & = \eta(\textbf{u}) \cdot \mu(t) \cdot \exp \left( \xi(\textbf{u},t) \right), \\
    \xi(\textbf{u},t | \theta , \alpha) & \sim \text{Gaussian process } (\beta,\phi(h_1,h_2|\theta , \alpha)), \\
    \alpha = (\alpha_t,\alpha_u,\alpha_e) & \sim \text{Smoothing parameters}, \\
    \theta & \sim \text{priors}
\end{split}
\label{ecu:inference}
\end{align}
where $\phi$ is the covariance function that comes with the solution of the spatio-temporal SPDE defined by (\ref{ecu:diffusion}). 

To fit the model, we use the \texttt{INLAspacetime} package \cite{lindgren2023inlaspacetime} by employing a finite basis function defined on a triangulated mesh of the region of study to approximate the spatio-temporal SPDE model described in (\ref{ecu:diffusion}) in this mesh \cite{simpson_et_al_2015}. 

As discussed in Section \ref{sec:spatemp}, the smoothing parameters regulate the smoothness of the Gaussian process in both space and time, as well as its type of separability. In summary, Table \ref{tab:diffmodels} presents the spatio-temporal models available in the \texttt{INLAspacetime} package, their type of separability, and the corresponding temporal and spatial smoothness parameters.

\begin{table}[h]
\centering
\begin{tabular}{ccccccc}
\hline Model & $\alpha_t$ & $\alpha_s$ & $\alpha_e$ & Type & $v_t$ & $v_s$ \\
\hline A & 1 & 0 & 2 & Separable order 1 & $1 / 2$ & 1 \\
B & 1 & 2 & 1 & Critical diffusion & $1 / 2$ & 1 \\
C & 2 & 0 & 2 & Separable order 2 & $3 / 2$ & 1 \\
D & 2 & 2 & 0 & Iterated diffusion & 1 & 2 \\
\hline
\end{tabular}
\caption{Models available in the R-package \texttt{INLAspacetime}.}
\label{tab:diffmodels}
\end{table}

In Table \ref{tab:diffmodels}, Model B (Critical diffusion) is nonseparable with a separability coefficient of $\beta_u = 0.5$. At the same time, Model D (Iterated diffusion) is fully non-separable with $\beta_u = 0$. Since velocity estimation requires computing first derivatives in both space and time, the intensity function of the LGCP process must be at least once differentiable. We focus on Models C and D to satisfy this requirement, ensuring the necessary smoothness. In Section \ref{sec:sim}, we evaluate and compare their performance.

\subsection{Velocities}
\label{sec:velocity}

With the intensity function of the underlying LGCP that governs an infectious disease, we can compute the \textit{minimum velocity}. This velocity comes from the advection-diffuse equation used to model the evolution of an implicit function $\psi(\textbf{u},t)$ in a specific domain. That is,

\begin{equation}
\label{ecu:adv_dif_1}
    \frac{\partial \psi(\textbf{u},t)}{\partial t} + \mathcal{S}(\textbf{u},t) \cdot \nabla \psi(\textbf{u},t) = 0,
\end{equation}
where $\nabla \psi(\textbf{u},t)$ is the spatial gradient of the function $\psi$ and $\mathcal{S}$ is the velocity that controls the direction and speed of the function $\psi$. These functions in the two dimensional spatial case, $\textbf{u} = (x,y)$ can be written explicitly as

\begin{align*}
\nabla \psi(\textbf{u},t) = \begin{pmatrix}
           \displaystyle\frac{\partial \psi(\textbf{u},t)}{\partial x} \\
           \vspace{-0.2cm} \\
           \displaystyle\frac{\partial \psi(\textbf{u},t)}{\partial y} \\
         \end{pmatrix}, \hspace{1cm}
\mathcal{S}(\textbf{u},t) = \left( \mathcal{S}_x(\textbf{u},t),\mathcal{S}_y(\textbf{u},t) \right).
\end{align*}

The velocity component can also be decomposed as $\mathcal{S}(\textbf{u},t) = s_{\mathbf{v}}(\textbf{u},t) \mathbf{v}(\textbf{u},t)$ where $\textbf{v}$ is a unit vector that represents the direction of the velocity and $s_\textbf{v}$ is its magnitude in the direction $\textbf{v}$. Under these assumptions, (\ref{ecu:adv_dif_1}) can be rewritten as

\begin{equation}
\label{ecu:adv_dif_2}
    \frac{\partial \psi(\textbf{u},t)}{\partial t} + s_{\mathbf{v}}(\textbf{u},t) [ \mathbf{v}(\textbf{u},t) \cdot \nabla \psi(\textbf{u},t) ] = 0.
\end{equation}

Now, for a fixed velocity direction $\textbf{v}$ and for a known function $\psi$, we can solve (\ref{ecu:adv_dif_2}) for the magnitude $s_{\textbf{v}}$ that is always positive. Then

\begin{equation}
\label{ecu:vel_v}
    s_{\textbf{v}}(\textbf{u},t) = \frac{\left| \partial \psi(\mathbf{u},t) / \partial t \right|}{\left|\textbf{v}(\textbf{u},v) \cdot \nabla \psi(\textbf{u},t)\right| }.
\end{equation}

Notice that the denominator in (\ref{ecu:vel_v}) is the absolute value of the spatial directional derivative of the function $\psi$ in the direction $\textbf{v}$, i.e. $D_{\textbf{v}} \psi(\textbf{u},t) = \textbf{v}(\textbf{u},v) \cdot \nabla \psi(\textbf{u},t)$. Then, using Cauchy-Schwart inequality, we know that the denominator will be bounded by the norm of the spatial gradient, $\left|\textbf{v}(\textbf{u},v) \cdot \nabla \psi(\textbf{u},t)\right| \leq \norm{\nabla \psi(\textbf{u},t)}$. Therefore, the largest instantaneous change of the function $\psi$ in a neighborhood of $(\textbf{u},t)$ is $\norm{\nabla \psi (\textbf{u},t)}$ and is in the direction $\textbf{v}_{\textbf{max}} = \nabla \psi(\textbf{u},t) / \norm{\nabla \psi(\textbf{u},t)}$. As a result, and because the bounded value is the denominator, the direction $\textbf{v}_{\text{max}}$ also provides the direction of minimum velocity at $(\textbf{u},t)$ that now could be written as

\begin{equation}
\label{ecu:prev_vel}
    s_{\text{min}}(\textbf{u},t) =  \frac{ \left| \partial \psi(\mathbf{u},t) / \partial t \right|}{\norm{\nabla \psi(\textbf{u},t)}}.
\end{equation}

In the point process context, following a similar procedure for the intensity function $\lambda(\textbf{u},t)$ of a spatio-temporal point process, the direction of minimum velocity captures the slowest speed of change in intensity in a product neighborhood of $(\textbf{u},t)$. That is, the velocity that provides the slowest speed of change in chance for an event at $(\textbf{u},t)$ is 

\begin{equation}
\label{ecu:vel}
    s_{\text{min}}(\textbf{u},t) =  \frac{\left| \partial \lambda(\mathbf{u},t) / \partial t \right|}{\norm{\nabla \lambda(\textbf{u},t)}}.
\end{equation}

The direction of this minimal velocity i.e., the direction of the slowest rate of intensity change, which remains positive i.e.d.

\begin{equation}
\label{ecu:dir_vel}
\vec{s}_{\text{min}}(\textbf{u},t) = \text{sign} \left( \frac{\partial \lambda(\textbf{u},t)}{\partial t} \right) \cdot \frac{\nabla \lambda(\textbf{u},t)}{\norm{\nabla \lambda(\textbf{u},t)}}.
\end{equation}
where the sign is a function that is $-1$ if the number is negative, $0$ if it is zero, or $+1$ if it is positive. This vector captures both the magnitude and direction of the slowest rate of change of intensity, making it a useful quantity for interpreting in the context of disease spread, how and where events are emerging or disappearing in space-time.

\subsection{Using velocities to understand disease spread}
\label{sec:velocity-exa}

\review{In this section, we present illustrative examples of intensity functions that evolve over time, along with their corresponding velocity fields. These examples help clarify how to interpret both the magnitude and direction of velocity in the context of infectious disease dynamics. Specifically, we consider three scenarios: one in which a hotspot emerges, another in which it disperses, and a third in which the hotspot shifts its location horizontally. All of the examples are defined spatially in the unit square $[0,1]^2$ and temporally in the unit interval $[0,1]$, and all the velocities are computed at the middle point at $t=0.5$. Also, the examples have the LGCP structure, so it is the exponential of a linear combination of the density of a bivariate normal distribution with a diagonal covariance matrix.}

\begin{figure}[!h]
    \centering
    \begin{subfigure}{0.32\textwidth}
        \begin{minipage}{1\textwidth}
            \centering
            \includegraphics[width=\linewidth]{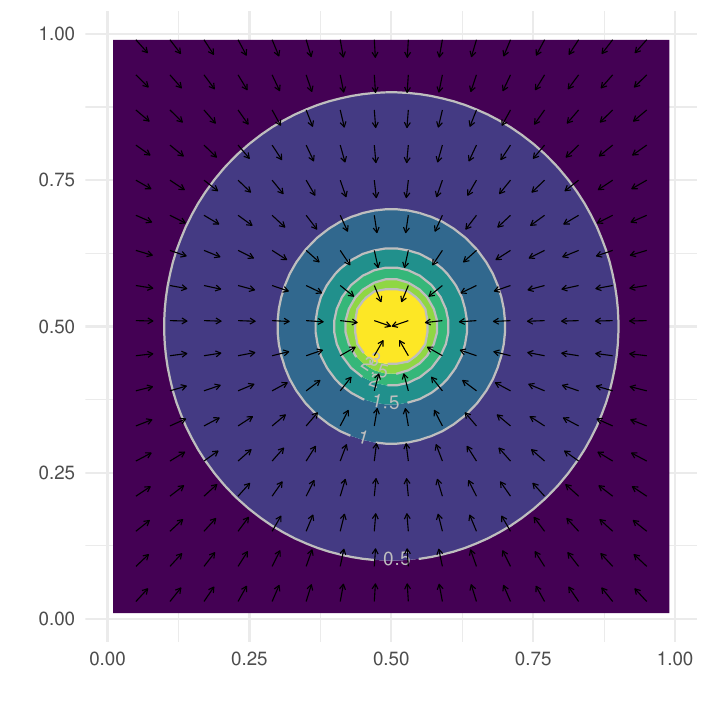}
            \caption{Example 1}
        \end{minipage}
    \end{subfigure}
    \begin{subfigure}{0.32\textwidth}
        \begin{minipage}{1\textwidth}
            \centering
            \includegraphics[width=\linewidth]{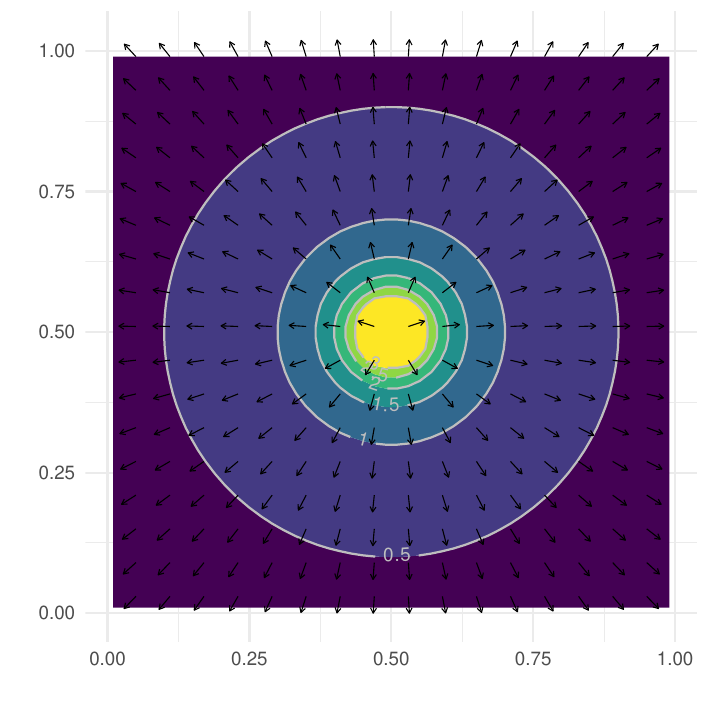}
            \caption{Example 2}
        \end{minipage}
    \end{subfigure}
    \begin{subfigure}{0.32\textwidth}
        \begin{minipage}{1\textwidth}
            \centering
            \includegraphics[width=\linewidth]{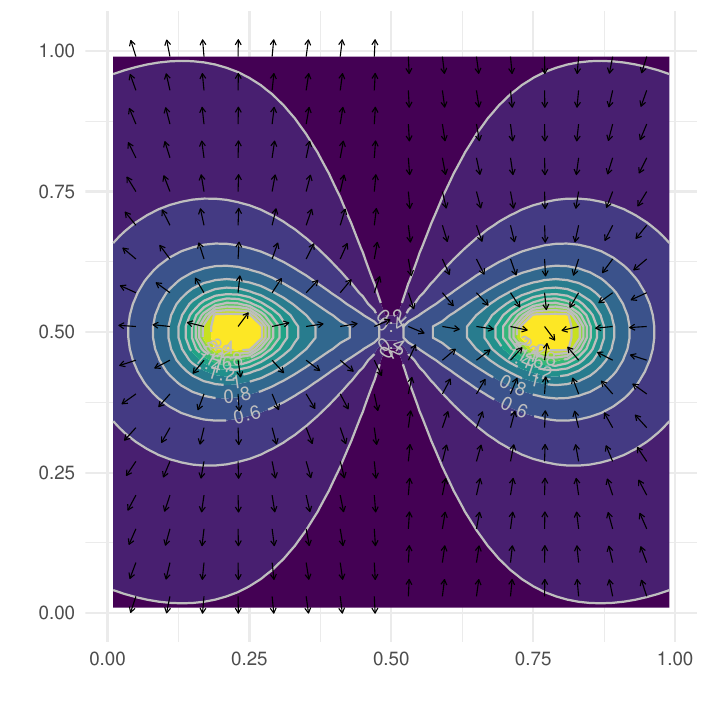}
            \caption{Example 3}
        \end{minipage}
    \end{subfigure}
    \caption{\review{Magnitude and direction of the true minimal velocities at $t = 0.5$. Subfigure (a) illustrates the emergence of a hotspot, (b) shows the disappearance of a hotspot, and (c) depicts the horizontal movement of a hotspot.}}
    \label{fig:vel_est_exa}
\end{figure}

\review{In the first scenario, the intensity is zero at time $t = 0$ and gradually increases over time. For each $t > 0$, the intensity has a single mode centered at $(0.5, 0.5)$, reaching its maximum at $t = 1$. This setup models the emergence of a central hotspot. As shown in Figure~\ref{fig:vel_est_exa} (a), at $t = 0.5$, the velocity is strongest in the center of the domain, with arrows pointing inward toward the hotspot. This inward flow produces a plughole-like vortex effect, indicating the formation and growth of a localized outbreak.

In the second scenario, the intensity pattern is reversed: it starts with a central hotspot at $(0.5, 0.5)$ at $t = 0$ with maximum intensity, which gradually decreases until it vanishes at $t = 1$. This represents a scenario where an initially concentrated outbreak becomes controlled and dissipates over time. As seen in Figure~\ref{fig:vel_est_exa} (b), the velocity magnitude is again highest in the central region; however, unlike the first case, the arrows point outward from the hotspot, producing a seed-dispersal-like effect that reflects the spatial diffusion or dissolution of the outbreak.

In the third scenario, a hotspot initially located at $(0.2, 0.5)$ at $t = 0$ shifts horizontally to $(0.8, 0.5)$ by $t = 1$. The mode remains strong throughout, but its spatial location changes. This setup models a situation where the outbreak relocates from one part of the region to another, indicating spatial movement of disease activity. In Figure~\ref{fig:vel_est_exa} (c), the velocity field shows two prominent zones: one on the left and one on the right. The left-hand hotspot displays outward-pointing arrows, signaling the disintegration of the hotspot. In contrast, the right-hand hotspot has inward-pointing arrows, indicating the accumulation of cases and emergence of a new hotspot.

Notice that in more complex scenarios, when there are more than two hotspots, there could also be scenarios of dispersion in zones of high velocity magnitude and arrows pointing to the same direction, as shown in the simulation study for $t=0.575$ in Figure \ref{fig:sim_vel} in the second row, first column. }

\subsection{Computing velocities}
\label{sec:velocity-est}

We estimate the minimal velocity $s_{\text{min}}(\textbf{u},t)$ using the output of the model described in Section \ref{sec:inference} which provides the estimated intensity function $\hat{\lambda}(\textbf{u},t)$ over a spatio-temporal grid. 
Let $\Delta_t$, $\Delta_x$, $\Delta_y$ denote the discrete step size in time $t$, latitude $x$ and longitude $y$ respectively, and let $\hat{\lambda}(i,j,n)$ be the estimated intensity function in the grid indexed in space for $(i,j)$ and time $n$. Then, the derivative of the intensity in time that appears in the numerator of $s_{\text{min}}(\textbf{u},t)$ can be approximated using numerical methods, such as \textit{backward differentiation} 
\begin{equation}
\label{ecu:apo_time}
    \frac{\partial \hat{\lambda}(\textbf{u},t)}{\partial t} \approx \frac{\hat{\lambda}(i,j,n)-\hat{\lambda}(i,j,n-1)}{\Delta_t}.
\end{equation}
Notice that the temporal derivative approximation presented in (\ref{ecu:apo_time}) needs future values in time. However, if we are interested in predictions, we can use a first-order divided difference if we do not have access to this value. 

To estimate the denominator $\norm{\nabla \lambda(\textbf{u},t)}$, following \cite{sethian1999}, we first define the forward and backward difference operators in space as follows

\begin{align*}
    & D^{+x} \hat{\lambda}(i,j,n) = \frac{\hat{\lambda}(i+1,j,n)-\hat{\lambda}(i,j,n)}{\Delta_x}, \\
    & D^{-x} \hat{\lambda}(i,j,n) = \frac{\hat{\lambda}(i,j,n)-\hat{\lambda}(i-1,j,n)}{\Delta_x}, \\
    & D^{+y} \hat{\lambda}(i,j,n) = \frac{\hat{\lambda}(i,j+1,n)-\hat{\lambda}(i,j,n)}{\Delta_y}, \\
    & D^{-y} \hat{\lambda}(i,j,n) = \frac{\hat{\lambda}(i,j,n)-\hat{\lambda}(i,j-1,n)}{\Delta_y} .
\end{align*}
Here, the positive exponent represents the forward differencing, and the negative represents the backward differencing. Then, using these values, we define the following differential operators in space

\begin{align*}
    & \norm{\nabla_{1} \hat{\lambda}(i,j,n)} = \left( (D^{+x} \hat{\lambda}(i,j,n))^2 + (D^{+y} \hat{\lambda}(i,j,n))^2 \right)^{1/2} ,\\
    & \norm{\nabla_{2} \hat{\lambda}(i,j,n)} = \left( (D^{+x} \hat{\lambda}(i,j,n))^2 + (D^{-y} \hat{\lambda}(i,j,n))^2 \right)^{1/2} ,\\
    & \norm{\nabla_{3} \hat{\lambda}(i,j,n)} = \left( (D^{-x} \hat{\lambda}(i,j,n))^2 + (D^{+y} \hat{\lambda}(i,j,n))^2 \right)^{1/2} ,\\
    & \norm{\nabla_{4} \hat{\lambda}(i,j,n)} = \left( (D^{-x} \hat{\lambda}(i,j,n))^2 + (D^{-y} \hat{\lambda}(i,j,n))^2 \right)^{1/2} .
\end{align*}
An approximation of the norm of the spatial gradient is then obtained as the average of the last four expressions
\begin{equation}
\label{ecu:apo_spa}
    \norm{\nabla \hat{\lambda} (\textbf{u},t)} \approx \left(\frac{1}{4} \right) \sum_{k=1}^{4} \norm{\nabla_{k} \hat{\lambda}(i,j,n)}.
\end{equation}

Finally, the minimal velocity described in Equation (\ref{ecu:vel}) can be approximated
using the approximations described in (\ref{ecu:apo_spa}) and (\ref{ecu:apo_time}), as follows
\begin{equation}
\label{ecu:aprox_vel}
    s_{\text{min}}(\textbf{u},t) \approx \frac{ \left| \displaystyle\frac{\hat{\lambda}(i,j,n)-\hat{\lambda}(i,j,n-1)}{\Delta_t} \right|}{\left(\frac{1}{4} \right) \displaystyle\sum_{k=1}^{4} \norm{\nabla_{k} \hat{\lambda}(i,j,n)}}.
\end{equation}

\section{Simulation study}
\label{sec:sim}

In this section, we perform a simulation study to assess the performance of our model in comparison with an underlying truth intensity function. Specifically, we consider a deterministic intensity function to generate a spatio-temporal point pattern in the unit cube, $(\textbf{u},t) \in [0,1]^3 $, defined by
\begin{equation}
\label{ecu:inten_sim}
    \lambda(\textbf{u},t) = \lambda_0 \exp \{ \beta_0 + (1-t) \beta_1 f_1(\textbf{u}) + t(1-t) \beta_2 f_2(\textbf{u}) + t^2 \beta_3 f_3(\textbf{u})\},
\end{equation}
where each $f_k(\textbf{u})$ denotes a bivariate normal density with mean $\boldsymbol{\mu}_k$ and covariance matrix $\Sigma_k$ for $k=1,2,3$, and $\lambda_0>0$ is a scaling parameter.

This intensity function, proposed in \cite{schliep_gelfand_2019}, was chosen because its close form allows to compute partial derivatives in space and time, and with that, we can compute the true minimal velocity described in (\ref{ecu:vel}).

The partial derivatives of this intensity function have the following form
\begin{align}
\begin{split}
    \frac{\partial \lambda(x,y,t)}{\partial x} & = \lambda(x,y,t) \left[ (1-t) \beta_1 f_{1 x}(x,y) + t(1-t) \beta_2 f_{2 x}(x,y) + t^2 \beta_3 f_{3 x}\right], \\
    \frac{\partial \lambda(x,y,t)}{\partial y} & = \lambda(x,y,t) \left[ (1-t) \beta_1 f_{1 y}(x,y) + t(1-t) \beta_2 f_{2 y}(x,y) + t^2 \beta_3 f_{3 y}\right], \\
    \frac{\partial \lambda(x,y,t)}{\partial t} & = \lambda(x,y,t) \left[ - \beta_1 f_{1}(x,y) + (1-2t) \beta_2 f_{2}(x,y) + 2t \beta_3 f_{3}\right],
\end{split}
\label{ecu:true-vel-1}
\end{align}
where for $k=1,2,3$, $f_{kx}$ denotes the partial derivative in the $x$ direction of $f_k$, and have the following closed form
\begin{align}
\begin{split}
    f_{k x}(x,y) & = \frac{\partial f_{k}(x,y)}{\partial x} \\
    & = f_{k}(x,y) \left( -(x-\mu_{k1}) \Sigma_{k,11}^{-1} - (y-\mu_{k 2}) \Sigma_{k,12}^{-1} \right),
\end{split}
\label{ecu:true-vel-2}
\end{align}
and $\Sigma_{k,ij}^{-1}$ is the $i$th row and $j$th column of $\Sigma_{k}^{-1}$. Analogously, the partial derivative of the density function $f_{k}$ in direction $y$ is
\begin{align}
\begin{split}
    f_{k y}(x,y) & = \frac{\partial f_{k}(x,y)}{\partial y} \\
    & = f_{k}(x,y) \left( -(x-\mu_{k1}) \Sigma_{k,12}^{-1} - (y-\mu_{k 2}) \Sigma_{k,22}^{-1} \right).
\end{split}
\label{ecu:true-vel-3}
\end{align}
To simulate the spatio-temporal point pattern, we fixed the parameters as $\beta_0=-1.5$, $\beta_1 = 8.0$, $\beta_2 = 2.0$, $\beta_3 = 2.0$, $\mu_1 = (0.4,0.2)$, $\mu_2 = (0.8,0.5)$, $\mu_3 = (0.2,0.8)$, and
\begin{equation*}
\Sigma_1=\left(\begin{array}{rr}0.065 & -0.030 \\ -0.030 & 0.065\end{array}\right),
\Sigma_2=\left(\begin{array}{ll}0.065 & 0.000 \\ 0.000 & 0.065\end{array}\right),
\Sigma_3=\left(\begin{array}{ll}0.065  & 0.030 \\ 0.030 & 0.065\end{array}\right).
\end{equation*}
Once we fixed the other parameters, for $\lambda_0$, which controls the number of points in the simulation, we proposed the following scenarios: $\lambda_0 \in [5,10,20,30]$.

With these fixed parameters, we use the \texttt{R} package \texttt{spatstat} \cite{baddeley_2015} to generate realizations of an inhomogeneous Poisson process with intensity function given by (\ref{ecu:inten_sim}) for each $\lambda_0$. This results in 359, 773, 1526, and 2276 points distributed on a grid with a temporal resolution of $1/20$ and a spatial resolution of $1/40$. These settings were chosen to capture spatio-temporal variability in the intensity function. Specifically, the process begins with a mode at $\mu_1$, then shifts from $\mu_1$ to $\mu_2$ for $t \in (0,5)$, moves from $\mu_2$ to $\mu_3$ for $t \in (0,1)$, and finally stabilizes at $\mu_3$.  

\begin{figure}[h]
    \centering
    \begin{minipage}{0.32\textwidth}
        \centering
        \includegraphics[width=\linewidth]{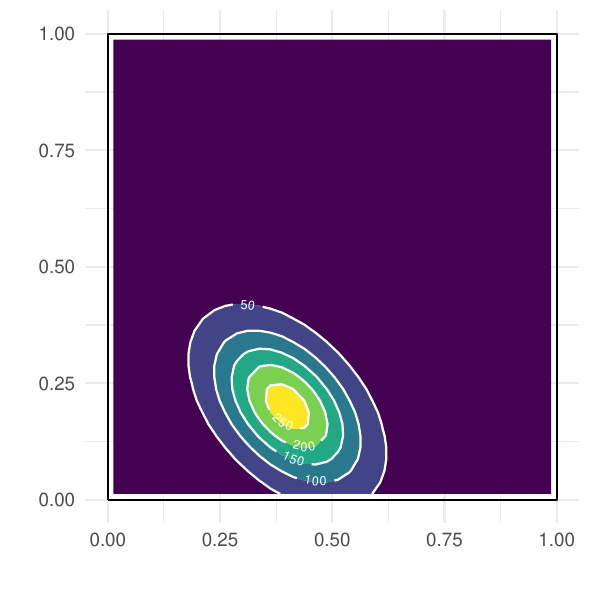}
        \caption*{Time = 0}
    \end{minipage}
    \begin{minipage}{0.32\textwidth}
        \centering
        \includegraphics[width=\linewidth]{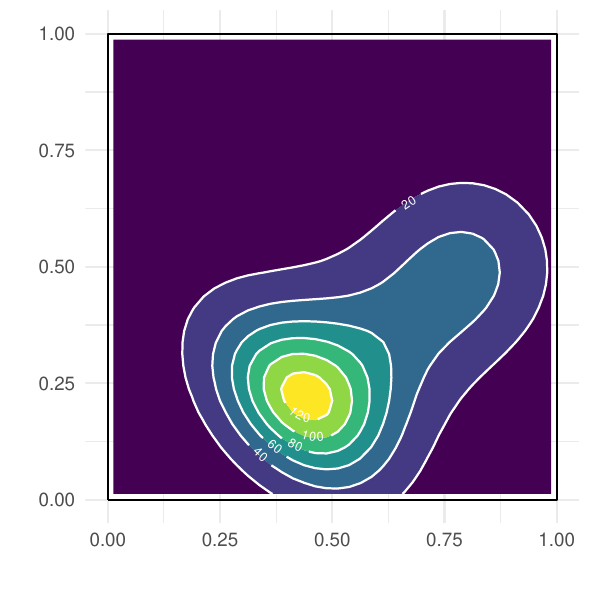}
        \caption*{Time = 0.25}
    \end{minipage}
    \begin{minipage}{0.32\textwidth}
        \centering
        \includegraphics[width=\linewidth]{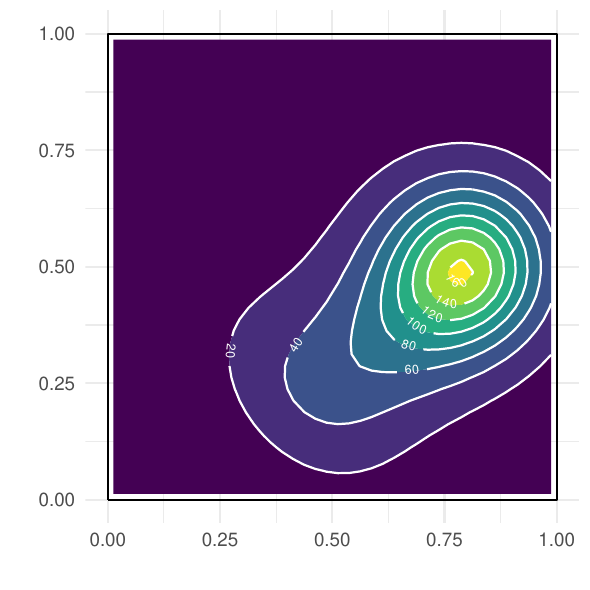}
        \caption*{Time = 0.5}
    \end{minipage}
    \begin{minipage}{0.32\textwidth}
        \centering
        \includegraphics[width=\linewidth]{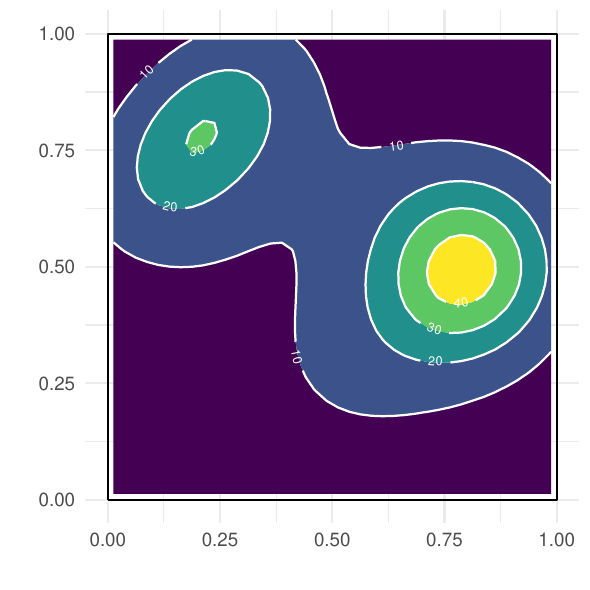}
        \caption*{Time = 0.75}
    \end{minipage}
    \begin{minipage}{0.32\textwidth}
        \centering
        \includegraphics[width=\linewidth]{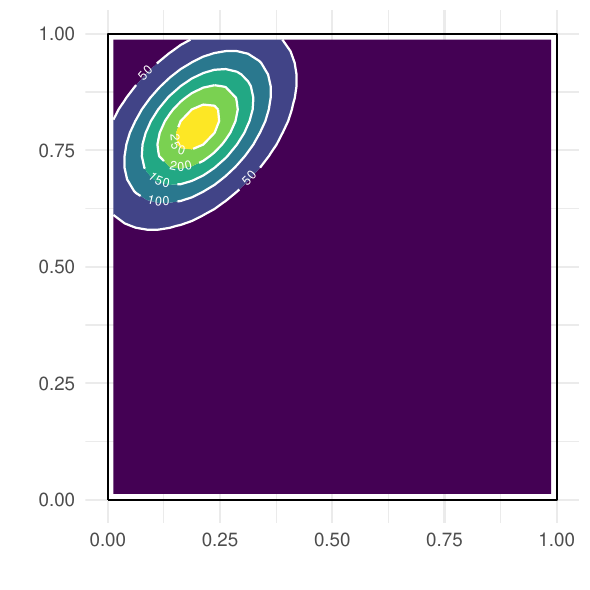}
        \caption*{Time = 1}
    \end{minipage}
    \caption{\review{Temporal evolution of the simulated intensity function at times $t = 0$, $0.25$, $0.5$, $0.75$, and $1$. Each panel shows the spatial distribution and variation of intensity over time, illustrating the movement of the mode over time.}}
    \label{fig:change_intensity}
\end{figure}

Figure \ref{fig:change_intensity} illustrates the intensity function for $\lambda_0 = 10$, showing how the mode evolves over time at points $0, 0.25, 0.5, 0.75$, and $1$. Note that $\lambda_0$ affects only the magnitude of the intensity function, leaving the mode unchanged across all cases.

\review{To illustrate the computation of velocity, we used three time points: $0.225$, $0.575$, and $0.875$. These were selected to capture key transition phases. First, $0.225$ marks the beginning of the change from the first mode $\mu_1$ to the second $\mu_2$, $0.575$ corresponds with the "middle" point of connections between the three modes, and $0.875$ represents the final shift from $\mu_2$ to $\mu_3$.}

Based on these times, Table \ref{tab:model_comparison} presents the root mean squared error (RMSE) between the velocity estimates obtained using (\ref{ecu:aprox_vel}) and the true values from  (\ref{ecu:true-vel-1}), (\ref{ecu:true-vel-2}), and (\ref{ecu:true-vel-3}) at times $0.225$, $0.575$, and $0.875$. The minimum RMSE for each case is highlighted in gray, comparing the LGCP model with a separable covariance structure (Model C) and the fully nonseparable model (Model D). The nonseparable model usually achieves a better fit, yielding a lower RMSE than its separable counterpart. This trend is particularly evident when the number of points is low ($\lambda_0=5$), where the fully nonseparable model is better than the separable model across all time points considered. This characteristic is especially relevant for epidemiological data, where case numbers are typically low in the early stages of disease spread. In such scenarios, the improved accuracy of the nonseparable model makes it a valuable tool for better capturing and analyzing disease dynamics.

\begin{table}[h]
\centering
\begin{tabular}{|c|c|c|c|}
    \hline
    Time & $\lambda_0$ & Model C & Model D \\
    \hline
    0.225 & 5  & 0.6959 & \cellcolor{lightgray}0.5154 \\
    \hline
    0.575 & 5  & 0.5866 & \cellcolor{lightgray}0.5302 \\
    \hline
    0.875 & 5  & 1.8021 & \cellcolor{lightgray}0.6843 \\
    \hline
    0.225 & 10 & 0.6216 & \cellcolor{lightgray}0.4917 \\
    \hline
    0.575 & 10 & \cellcolor{lightgray}0.5098 & 0.5511 \\
    \hline
    0.875 & 10 & \cellcolor{lightgray}0.8678 & 0.8721 \\
    \hline
    0.225 & 20 & 0.6366 & \cellcolor{lightgray}0.5754 \\
    \hline
    0.575 & 20 & \cellcolor{lightgray}0.5006 & 0.5654 \\
    \hline
    0.875 & 20 & 0.9694 & \cellcolor{lightgray}0.9552 \\
    \hline
    0.225 & 30 & 0.8499 & \cellcolor{lightgray}0.5480 \\
    \hline
    0.575 & 30 & 0.6003 & \cellcolor{lightgray}0.5273 \\
    \hline
    0.875 & 30 & \cellcolor{lightgray}0.7827 & 0.8578 \\
    \hline
\end{tabular}
\caption{\review{Root mean squared error (RMSE) of the estimated velocities for the separable model (Model C) and the non-separable model (Model D), compared to the true values. For each configuration, the lower RMSE between the two models is highlighted in grey.}}
\label{tab:model_comparison}
\end{table}

\review{In Appendix \ref{sec:appendix-modeling}, we present the estimation of the intensity function using \texttt{INLAspacetime} and \texttt{inlabru} for time points $0.225$, $0.575$, and $0.875$, with $\lambda_0 = 5$. We also describe the approximation of the absolute value of the temporal derivative and the norm of the spatial gradient of the intensity function, computed using finite differences as outlined in Section \ref{sec:velocity-est}. The temporal derivative is approximated using backward differences with a step size of $\Delta_t = 0.2$, while the spatial partial derivatives are calculated using both forward and backward differences with a step size of $\Delta_x = \Delta_y = 1/40$.}

\begin{figure}[!htp]
\captionsetup{skip=5pt}

\centering
\textbf{Velocity of the spatio-temporal point pattern}\par\medskip

\vspace{-0.3em}

\begin{minipage}{0.32\textwidth}
    \centering
    \textbf{True Values}
\end{minipage}
\begin{minipage}{0.32\textwidth}
    \centering
    \textbf{Model C}
\end{minipage}
\begin{minipage}{0.32\textwidth}
    \centering
    \textbf{Model D}
\end{minipage}

\vspace{0.3em}

\begin{subfigure}{1\textwidth}
    \captionsetup{skip=1pt}
    \begin{minipage}{0.32\textwidth}
        \centering
        \includegraphics[width=\linewidth]{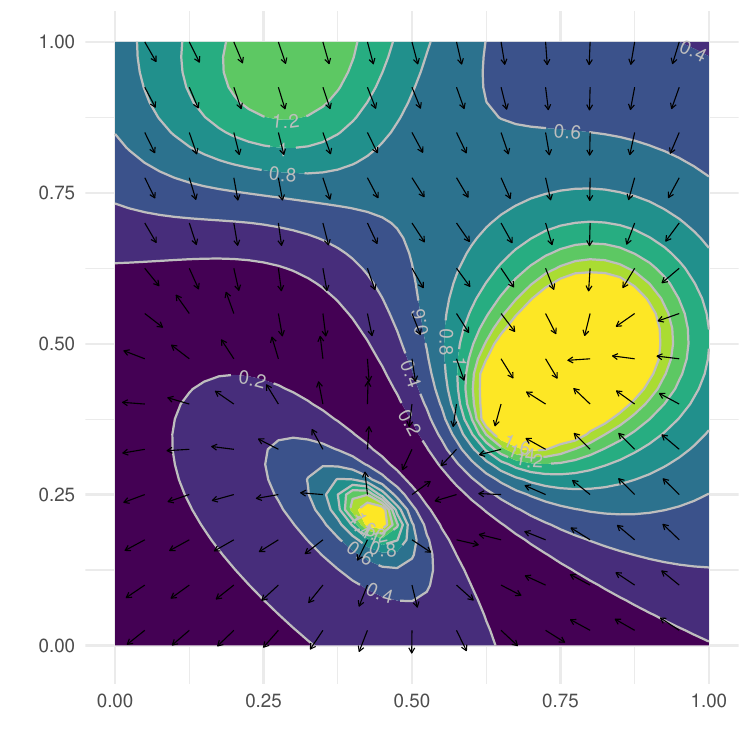}
    \end{minipage}
    \begin{minipage}{0.32\textwidth}
        \centering
        \includegraphics[width=\linewidth]{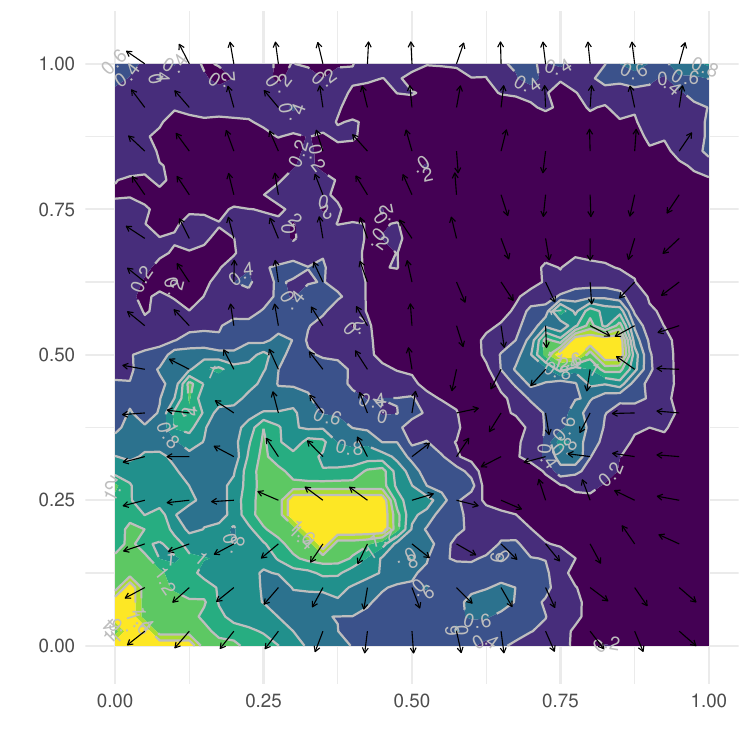}
    \end{minipage}
    \begin{minipage}{0.32\textwidth}
        \centering
        \includegraphics[width=\linewidth]{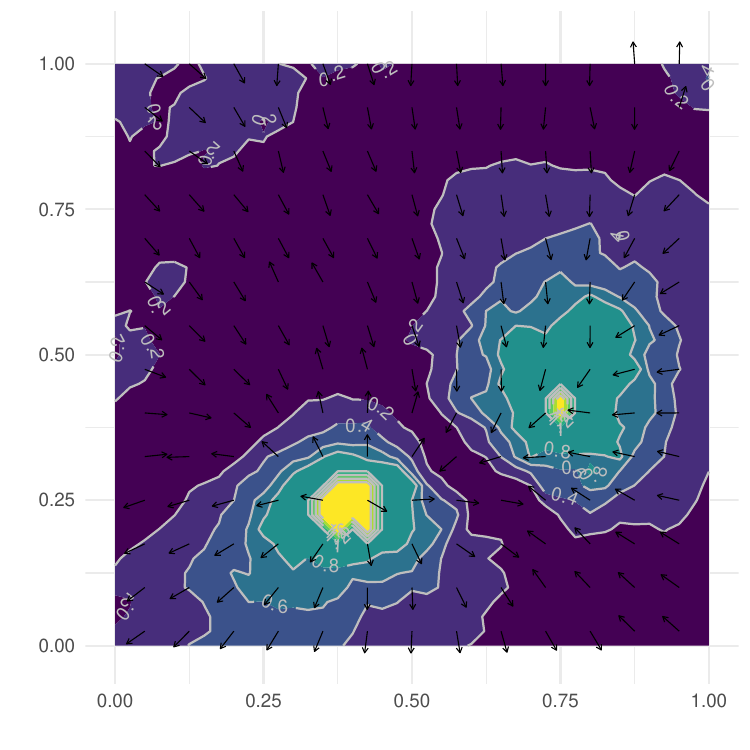}
    \end{minipage}
    \caption*{Time = 0.225}
\end{subfigure}

\vspace{-0.5em}

\begin{subfigure}{1\textwidth}
    \captionsetup{skip=1pt}
    \begin{minipage}{0.32\textwidth}
        \centering
        \includegraphics[width=\linewidth]{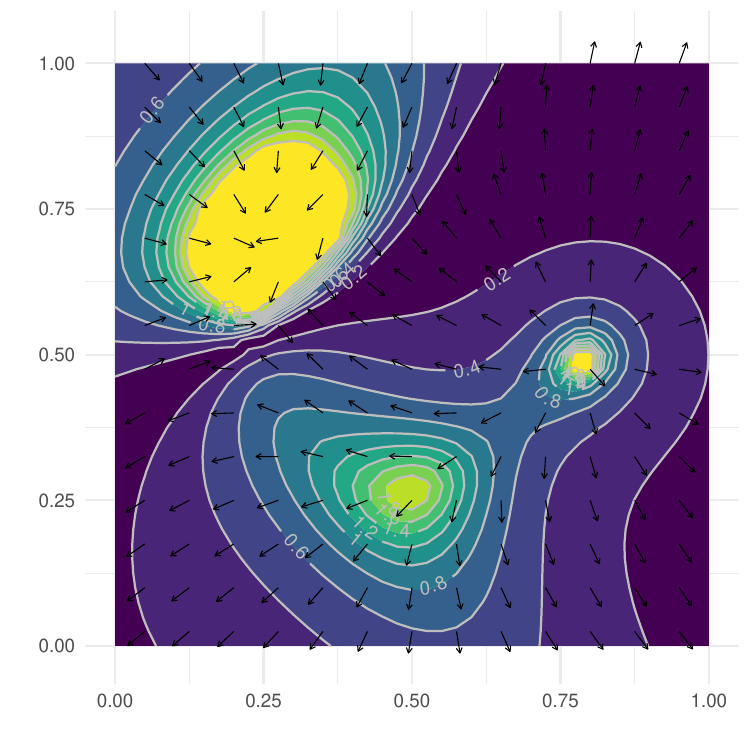}
    \end{minipage}
    \begin{minipage}{0.32\textwidth}
        \centering
        \includegraphics[width=\linewidth]{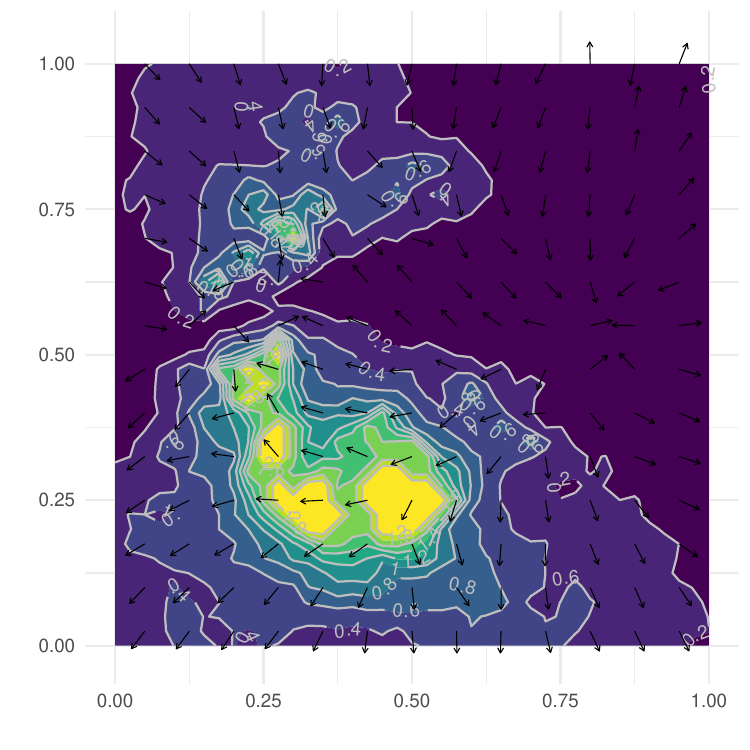}
    \end{minipage}
    \begin{minipage}{0.32\textwidth}
        \centering
        \includegraphics[width=\linewidth]{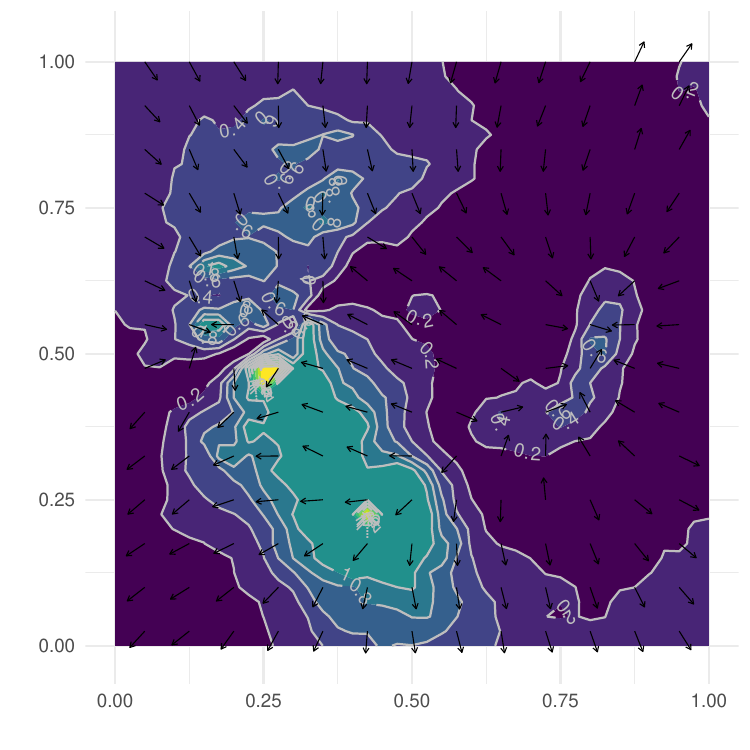}
    \end{minipage}
    \caption*{Time = 0.575}
\end{subfigure}

\vspace{-0.5em}

\begin{subfigure}{1\textwidth}
    \captionsetup{skip=1pt}
    \begin{minipage}{0.32\textwidth}
        \centering
        \includegraphics[width=\linewidth]{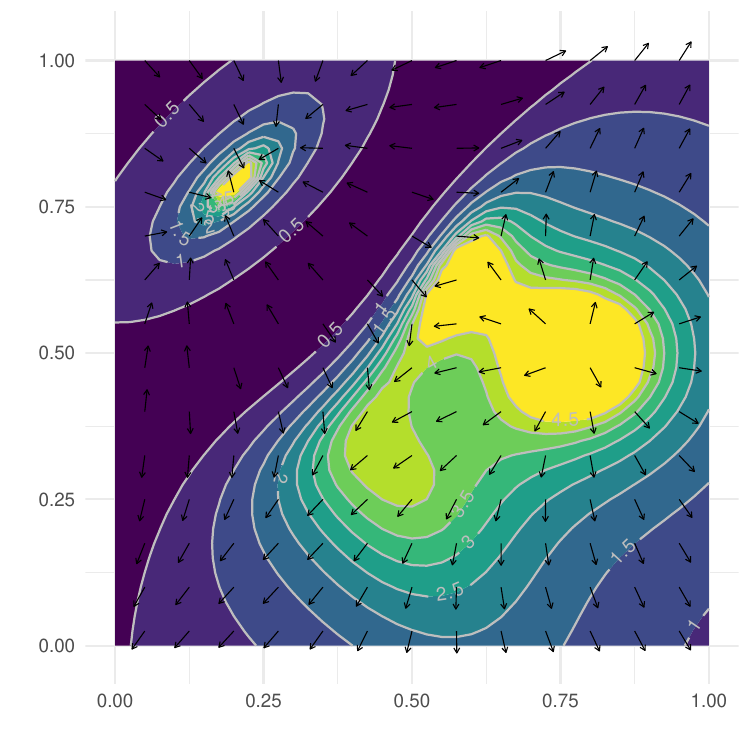}
    \end{minipage}
    \begin{minipage}{0.32\textwidth}
        \centering
        \includegraphics[width=\linewidth]{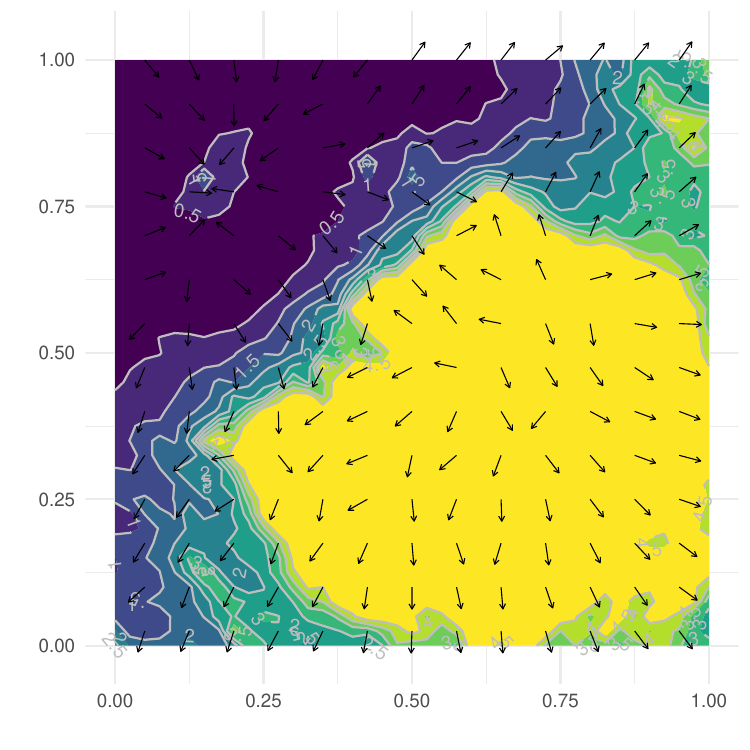}
    \end{minipage}
    \begin{minipage}{0.32\textwidth}
        \centering
        \includegraphics[width=\linewidth]{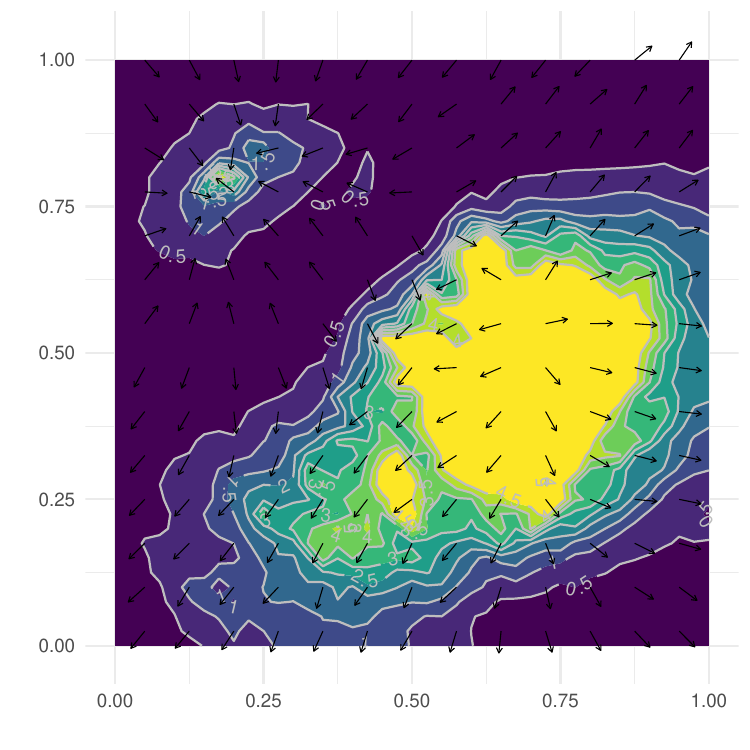}
    \end{minipage}
    \caption*{Time = 0.875}
\end{subfigure}

\caption{\review{Velocity estimation for the simulated spatio-temporal point pattern. Columns correspond to the true velocity (left), the estimate obtained using the separable model (Model C, center), and the estimate from the nonseparable model (Model D, right). Rows represent three key time points in the process: 0.225 (top), 0.575 (middle), and 0.875 (bottom). Each panel shows the magnitude and direction of the minimal velocity.}}
\label{fig:sim_vel}
\end{figure}

Figure \ref{fig:sim_vel} presents the true and estimated velocities at time points $0.225$, $0.575$, and $0.875$ for $\lambda_0=5$. The velocity estimates are obtained from the posterior mean of the diffusion model, where partial derivatives in time and space are computed and subsequently used in (\ref{ecu:aprox_vel}) to derive minimal velocities. The results indicate that the non-separable model (Model D) generally provides a more accurate velocity estimation compared to the Separable model (Model C) at time points $0.225$ and $0.875$. Additionally, the non-separable (Model D) yields a smoother velocity field across all time points, particularly at $t=0.575$, where it significantly reduces noise compared to the separable model (Model C). Finally, the arrows in Figure \ref{fig:sim_vel} are computed according to equation~(\ref{ecu:dir_vel}), where the sign of the temporal derivative determines orientation, and the gradient of the intensity defines the direction of steepest spatial change.

\review{All code used for this simulation study is publicly available at \url{https://github.com/fravellaneda/estimating-velocities}.}

\section{COVID-19 spread and velocities in Colombia}
\label{sec:app}

\begin{figure}[!htp]
\centering
\includegraphics[width=\textwidth]{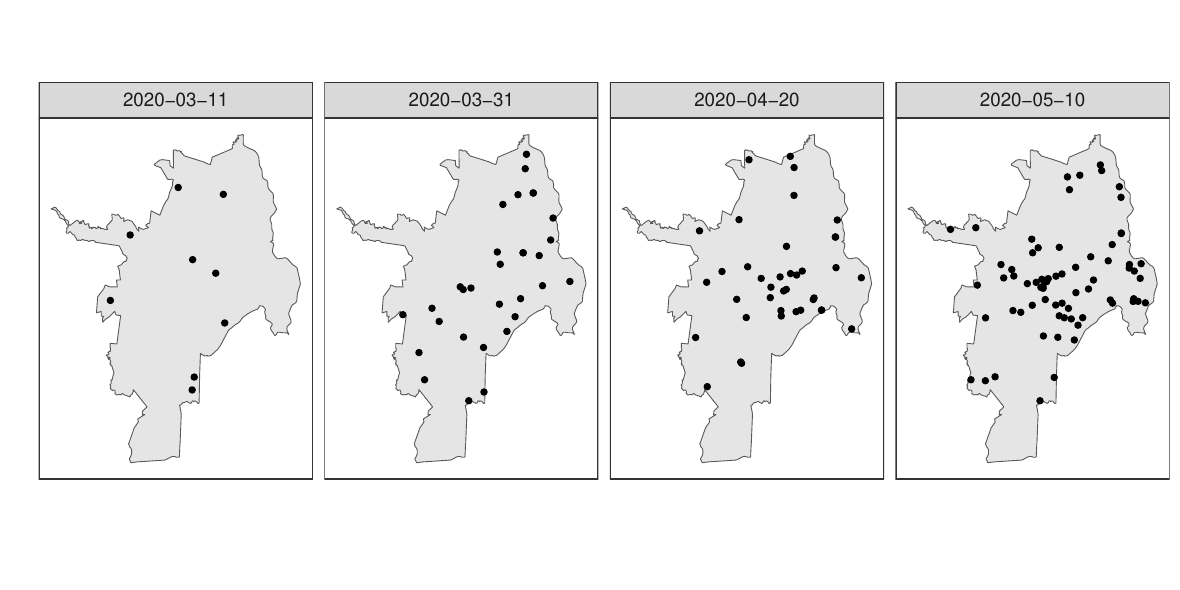}
\caption{\review{Spatial distribution of reported COVID-19 cases in Cali, Colombia, on four selected dates: March 11, March 31, April 20, and May 10, 2020. Each point represents the household location of an infected individual.}}
\label{fig:spatio-temp-covid}
\end{figure}

In this section, we applied our method to compute minimal velocities to understand the spread of an infectious disease. In particular, we use the household locations of COVID-19-infected individuals in Cali, one of the largest and most populated cities in Colombia, during the COVID-19 pandemic in 2020. Data were provided by the Cali Municipal Public Health Secretary \cite{Municipality_Cali} and have confirmed locations of 208551 infected individuals from 2020-03-01 to 2022-04-25. To begin, the geographical coordinates of the data were converted to UTM coordinates, and the values in the map are presented in meters. To model the LGCP,  we used the spatio-temporal SPDE described in (\ref{ecu:diffusion}) with the parameters given by model D. This process involves the construction of a triangulated mesh covering the study region, in this case, the city boundary of Cali, Colombia. Once we have fitted the model, we make predictions in a grid with cells equally spaced 1km apart, which results in 296 cells inside the city limits. For the temporal component, for simplicity, we use the first 80 days. In Figure \ref{fig:spatio-temp-covid}, we present the location of infected people on 4 different days in Cali, Colombia.

\subsection{Intensity function}

Following the ideas of Section \ref{sec:lgpc_general}, we apply a LGCP model with intensity function decomposed as the product of the spatial $\eta(\textbf{u})$, temporal $\mu(t)$, and a residual spatio-temporal variation $\exp \left( \xi (\textbf{u},t) \right)$ as follows
\begin{equation}
    \label{ecu:geninten2}
    \Lambda(\textbf{u},t)= \eta(\textbf{u}) \cdot \mu(t) \cdot \exp \left( \xi (\textbf{u},t) \right).
\end{equation}

In practice, this decomposition requires first computing the spatial and temporal components separately and then using this information to compute the spatio-temporal component. The computation of each component is detailed below.

\begin{figure}[!h]
\centering
\includegraphics[scale=0.85]{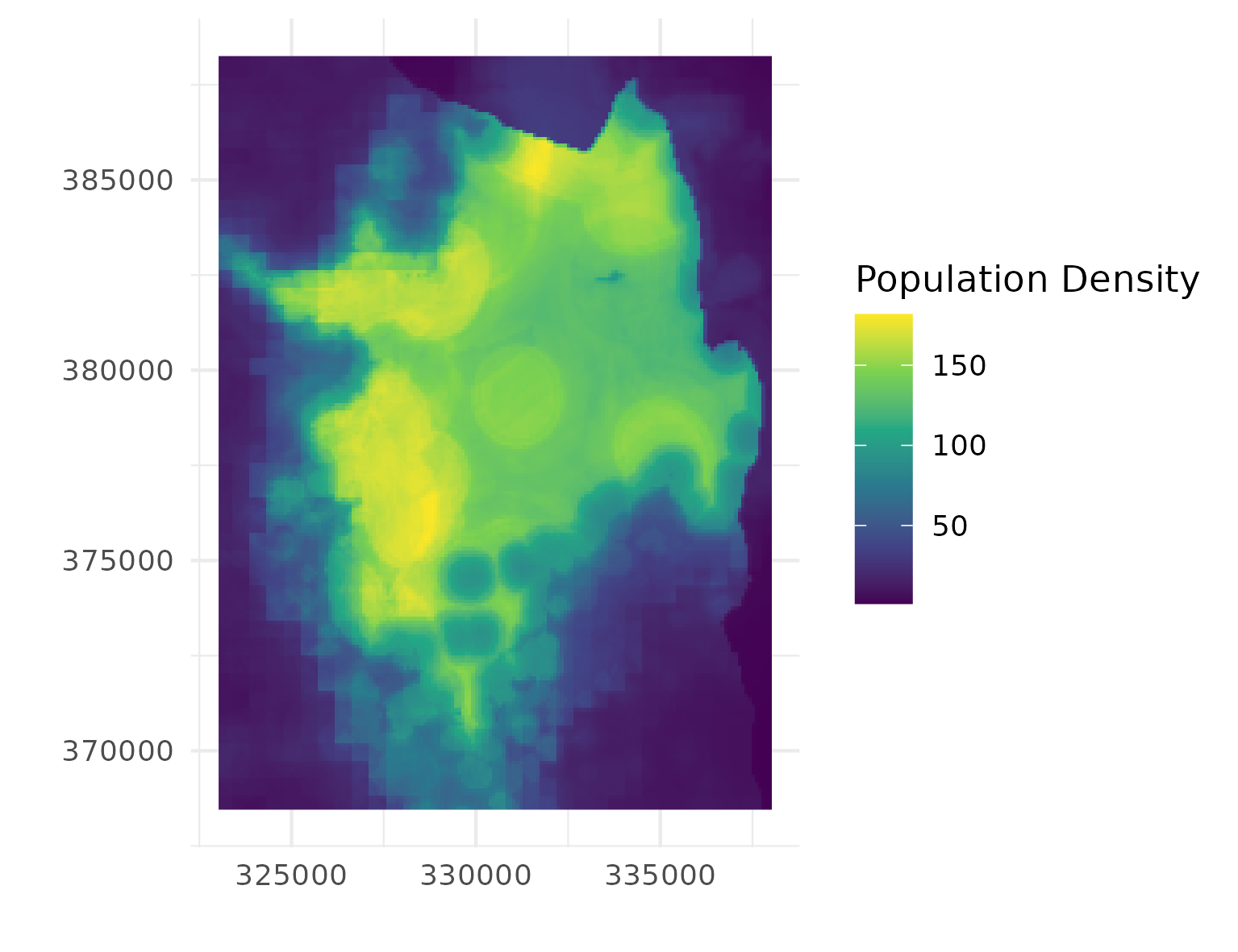}
\caption{\review{Estimated population density in Cali, Colombia, measured as the number of people in each cell ($100 m \times 100 m$).}}
\label{fig:pop_den}
\end{figure}

The spatial component represents the expected number of infected individuals in the city of Cali. To estimate this, we consider as a study region the city boundary of Cali. We use the \texttt{R-INLA} package to construct a triangulated mesh in this region, and compute the expected number of events using the dual mesh, which comprises a series of polygons surrounding each vertex of the original mesh.
On the other hand, we obtain the population density in Cali from WorldPop, an open-access tool for population data. In this case, we used a resolution of $100 \times 100$ $m$ cells defined in a square region that includes Cali (Figure \ref{fig:pop_den}).
Then, the expected number of events in each polygon of the dual mesh is computed as the area of the polygon multiplied by the average population of the cells in that polygon \cite{krainski_et_al_2019}.

\begin{figure}[!h]
\centering
\includegraphics[scale=0.6]{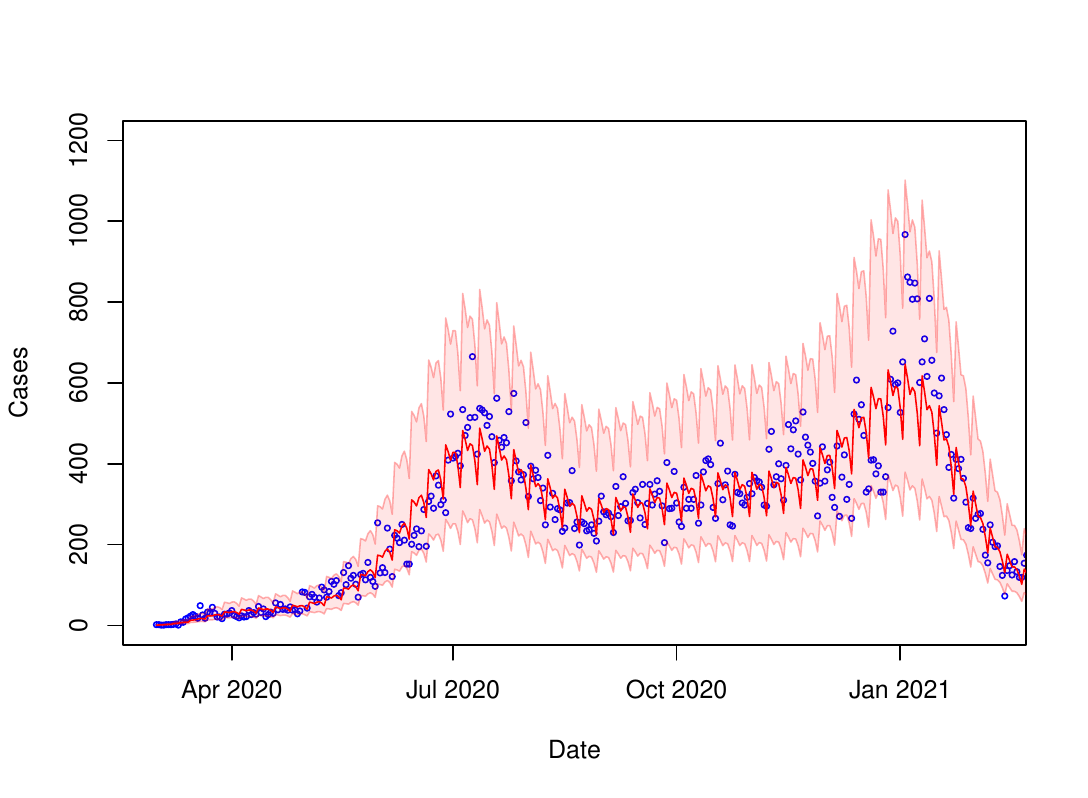}
\caption{\review{Daily COVID-19 cases (blue dots) with the fitted model (solid red line) and its associated prediction interval (shaded pink area) over time. The width of the prediction interval varies, ranging from a lower bound of $1.24$ to an upper bound of $264.3$.}}
\label{fig:timefit}
\end{figure}

To estimate the temporal component representing the daily expected number of infected people, we use a spline model as explained in Section \ref{sec:temp}. The selected basis is related to the structure of the data. It takes into account first the day-of-week effects, periodical effects described by sine and cosine functions called the Fourier basis, and finally an overall rising trend over time captured by a polynomial structure. The selected number of bases in each case was related to the simplicity and explainability of the model. In summary, the model has the following form

\begin{equation*}
    \log(\mu(t)) = \delta_{d_i} + \sum_{k=1}^{3} \left( \alpha_k \cos\zeta(k \omega t) + \beta_k \sin(k \omega t) + \gamma_k t^k \right),
\end{equation*}
where $\omega = 2 \pi/365$ is the annual periodicity. The estimators for the day-of-week effects are $\delta_d = \{ -4.52,-4.58,-4.64,-4.60,-4.62,-4.70,-4.86\}$, where $d_1$ is Sunday, $d_2$ is Monday, etc. In addition, the estimators of the Fourier basis are $\alpha = \{3.59,0.63,0.34\}$, and $\beta = \{-18.71,-2.46,-0.50\}$. Finally, estimators for the polynomial basis for degrees $k=1,2,3$ are $\gamma = \{0.69,-0.004,8.31e-06\}$. Figure \ref{fig:timefit} compares the result of the fitted model and the daily cases observed.

\begin{figure}[H]
\centering
\includegraphics[width=\textwidth]{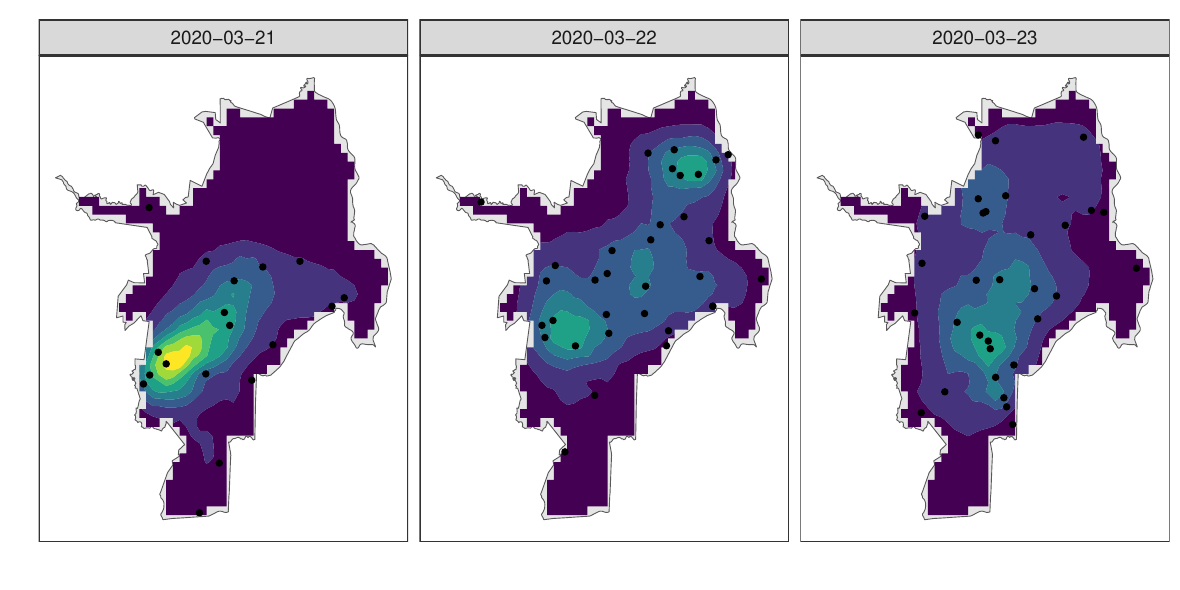}
\caption{\review{Posterior mean of the intensity function for March 21, 22, and 23, 2020. Black dots indicate the household locations of infected individuals on each respective day.}}
\label{fig:app_inten}
\end{figure}

Finally, using the spatial and temporal components previously estimated as an offset, we estimate the parameters of the spatio-temporal random effect  $\xi(\textbf{u},t)$ and obtain an estimate for the posterior distribution of the intensity function $\lambda(\textbf{u},t)$ using \texttt{INLAspacetime} and \texttt{inlabru} \cite{lindgren2023inlaspacetime,inlabru}. In Figure \ref{fig:app_inten}, we present the posterior mean of the intensity function for three different days. 

\subsection{Velocities}

\review{Using the estimated intensity function, we can compute the velocity and its direction within the spatial domain using  (\ref{ecu:aprox_vel}) and (\ref{ecu:dir_vel}). Figure \ref{fig:app_vel} illustrates these estimations for March 22, 2020 (a), and March 23, 2020 (b). Specifically, in Figure \ref{fig:app_vel} (a), the upper right region of the map exhibits two zones with high velocity magnitude. The arrows in the upper zone of these two are pointing towards the center, indicating the formation of a hotspot. This pattern suggests the emergence of new cases on March 22, 2020, in the upper right part of Cali, which were not present the previous day, as shown in Figure \ref{fig:app_inten}. For the second region, the lack of a clear arrow pattern prevents drawing definitive conclusions.

Additionally, Figure \ref{fig:app_vel} (a) also highlights a high value of the velocity magnitude presented in the left lower region. The arrows in this region, pointing to the left and right of the map, indicate that the number of infected individuals in that region is having a dispersal effect. That means that there was a significant number of cases on the day before that starts to disperse in that region, as it is shown in Figure \ref{fig:app_inten}.

\begin{figure}[!h]
    \centering

    \begin{subfigure}{0.47\textwidth}
        \centering
        \includegraphics[width=\linewidth]{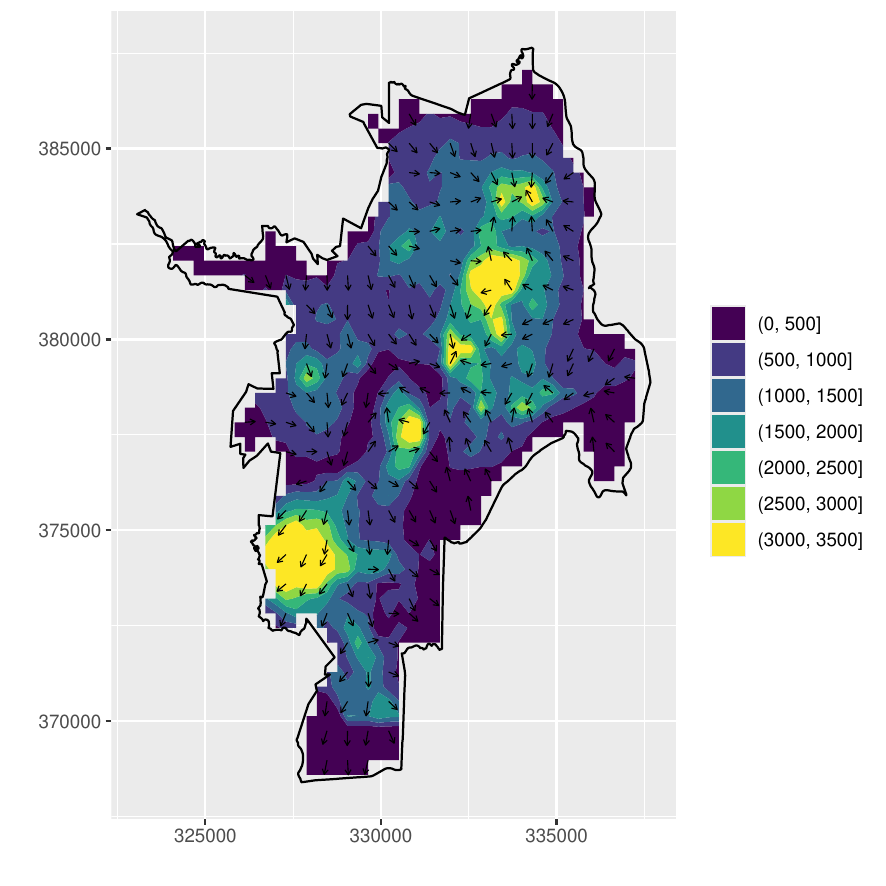}
        \caption{\review{Estimated velocity on March 22, 2020.}}
        \label{fig:vel_22_app}
    \end{subfigure}
    \hfill
    \begin{subfigure}{0.47\textwidth}
        \centering
        \includegraphics[width=\linewidth]{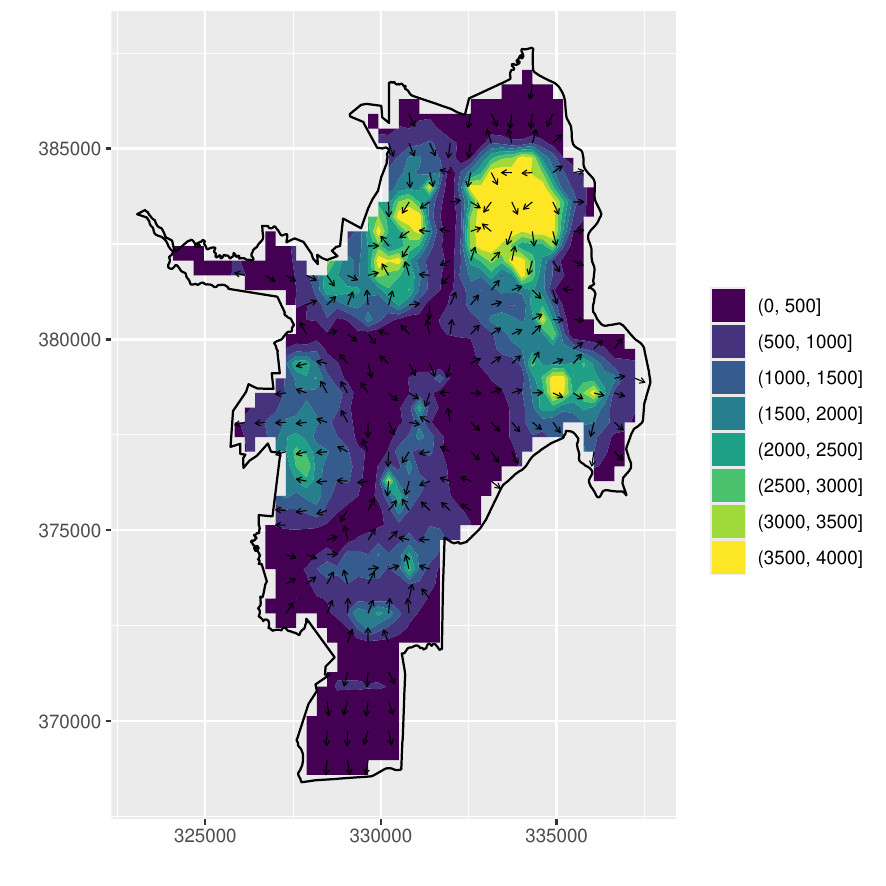}
        \caption{\review{Estimated velocity on March 23, 2020.}}
        \label{fig:vel_23_app}
    \end{subfigure}

    \caption{\review{Estimated minimal velocity $s_{\text{min}}(\mathbf{u}, t)$ and its direction for March 22 (left) and March 23 (right), 2020.}}
    \label{fig:app_vel}
\end{figure}

In Figure \ref{fig:app_vel}, the upper right region of the map exhibits two zones with a high velocity magnitude. The arrows in the left zone are pointing towards the center, indicating the formation of a hotspot. This pattern suggests the emergence of new cases on March 23, 2020, in the upper left part of Cali, which were not present the previous day, as shown in Figure \ref{fig:app_inten}. On the other hand, the zone on the right lacks a clear pattern, preventing any strong conclusions about its behavior.

We remind readers that this is a data-driven method, and data inaccuracies may affect the model’s accuracy. Such errors may introduce noise in either the magnitude or the direction of the velocity, though typically not in both simultaneously. Therefore, we strongly recommend drawing inferences only when both conditions, high magnitude and clear direction, are present, as the presence of only one is not sufficient.}

\section{Conclusions and discussion}
\label{sec:con}

In this paper, we present a spatio-temporal modeling approach
to estimate the velocities of infectious disease spread. We model the locations and times of people infected using a spatio-temporal log-Gaussian Cox point process, and obtain the velocities using finite differences that approximate the partial derivatives of the intensity function.

Specifically, we present an alternative derivation of the formula for the minimal velocity based on the advection-diffuse equation for a sufficiently smooth function. We use that derivation as a starting point to extend it to the intensity function of a spatio-temporal point process. We use an LGCP model and consider a multiplicative decomposition of the space-time intensity function into spatial, temporal, and residual spatio-temporal components. First, the spatial component representing the expected number of cases can be estimated using a nonparametric estimation by kernels or by integrating population data. Second, we proposed a spline model for the temporal component that uses day-of-week, periodic, and polynomial growing effects. Finally, for the spatio-temporal component, we used a Matérn covariance structure on space and an independent structure over time. For inference, we used \texttt{INLAspacetime} and used the estimation of the intensity function to propose an approximation of the minimal velocity using finite differences.

Through a simulation study, we demonstrated that this approach is effective, with the approximation by finite differences yielding an overall good fit. That happens because the point approximation of the intensity function is smooth enough, which is one of the advantages of approximating an LGCP using basis functions in \texttt{INLAspacetime}. One of the main contributions of the velocity approximation proposed here is that we do not use a parametric estimation of the spatial or temporal gradients. Here, we propose a model-free approximation of the velocity that is flexible enough to capture \textit{instantaneous} changes in the intensity function between two time steps, given the data's time resolution, for instance, the change of an infectious disease from one day to the next. We use first-order approximations of the partial derivatives in space and time mainly because of their simplicity. However, for future work, we can improve the approximation by using the smoothing properties of the intensity function to propose second-order approximations of the partial derivatives. 

\clearpage

\review{We demonstrated the effectiveness of our method by computing minimal velocities from COVID-19 data in Cali, Colombia, showing how these estimates help to understand disease spread. In this application, we showed  that velocity provides richer information about the dynamics of disease propagation than first-order estimators. This insight is especially useful when existing approaches fail to capture directional patterns. In addition, this work could be extended to model the velocities of multiple infectious diseases interacting in the same population, such as dengue and chikungunya, which are vector-borne diseases with common risk factors \cite{pavanietal23}. Our method can also be applied in other fields to study the direction and speed of movement in point patterns -- such as species, crimes, or fires -- to better understand how their geographic distribution changes over time and how it relates to climate or other factors \cite{moraga20,moraga23}.}

In summary, we presented a method for approximating the minimal velocity and direction of disease spread using finite differences of the intensity function, modeled with a spatio-temporal process. This is a data-driven approach that uses a flexible model that could be applied in infectious diseases or other spatio-temporal point patterns such as crime data or species locations, taking into account that the covariance structure for these problems is not the same as in contagious diseases. 

\clearpage

\bibliographystyle{apalike}
\bibliography{biblio.bib}       


\clearpage

\begin{appendices}

\section{Modeling results}
\label{sec:appendix-modeling}

\begin{figure}[htp]

\vspace{-0.5em}

\captionsetup{skip=5pt}
\centering
\textbf{Intensity function}\par\medskip

\vspace{-0.3em}

\begin{minipage}{0.32\textwidth}
    \centering
    \textbf{True Values}
\end{minipage}
\begin{minipage}{0.32\textwidth}
    \centering
    \textbf{Model C}
\end{minipage}
\begin{minipage}{0.32\textwidth}
    \centering
    \textbf{Model D}
\end{minipage}

\vspace{0.3em}

\begin{subfigure}{1\textwidth}
    \captionsetup{skip=1pt}
    \begin{minipage}{0.32\textwidth}
        \centering
        \includegraphics[width=\linewidth]{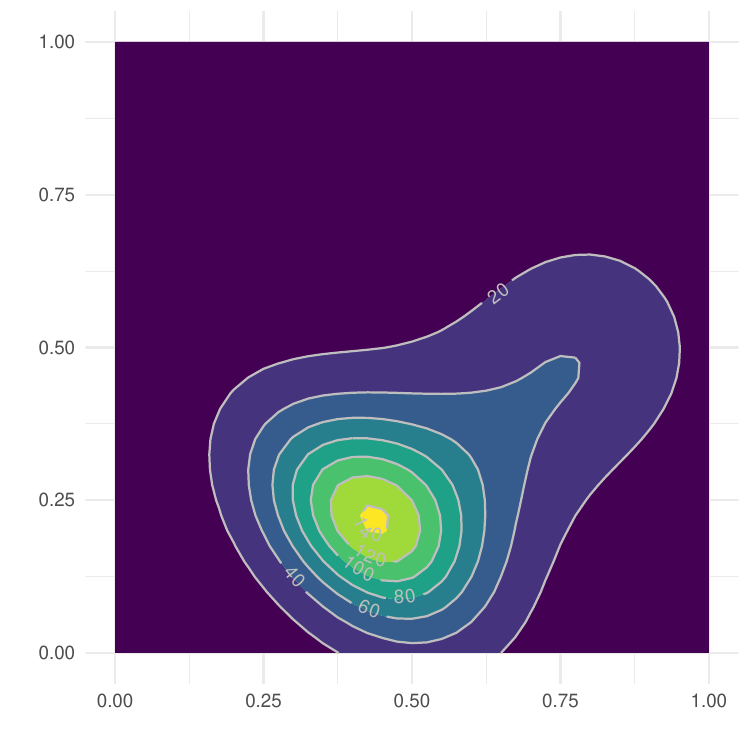}
    \end{minipage}
    \begin{minipage}{0.32\textwidth}
        \centering
        \includegraphics[width=\linewidth]{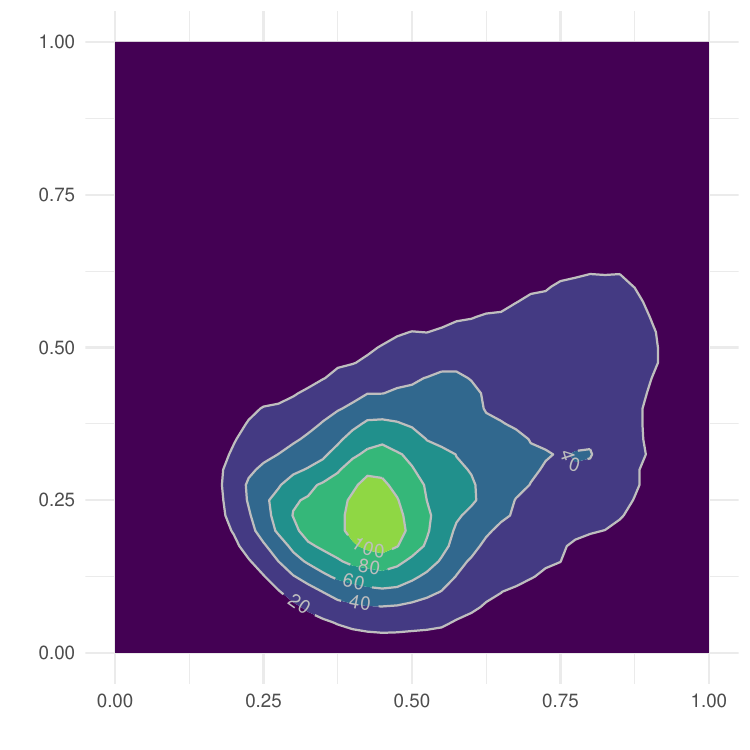}
    \end{minipage}
    \begin{minipage}{0.32\textwidth}
        \centering
        \includegraphics[width=\linewidth]{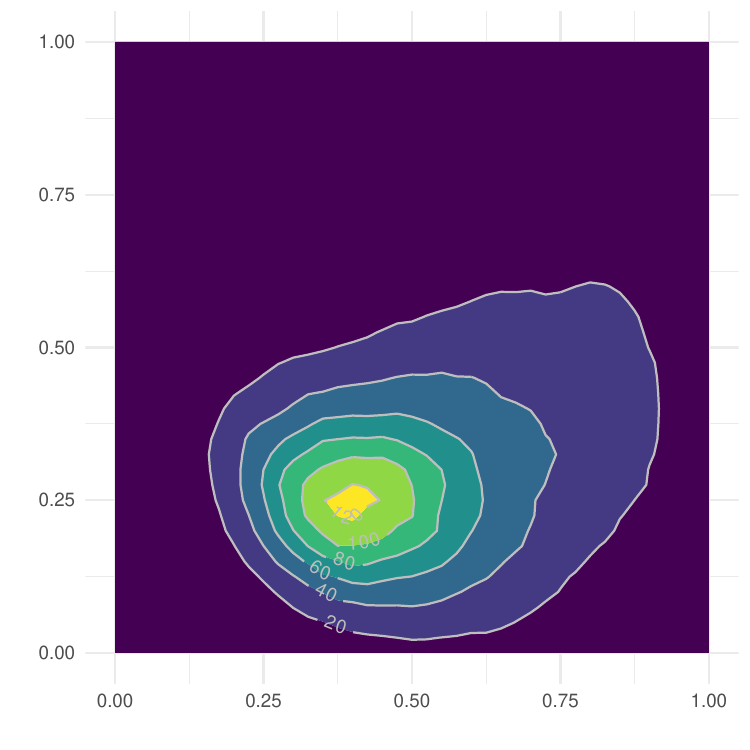}
    \end{minipage}
    \caption*{Time = 0.225}
\end{subfigure}

\vspace{-0.5em}

\begin{subfigure}{1\textwidth}
    \captionsetup{skip=1pt}
    \begin{minipage}{0.32\textwidth}
        \centering
        \includegraphics[width=\linewidth]{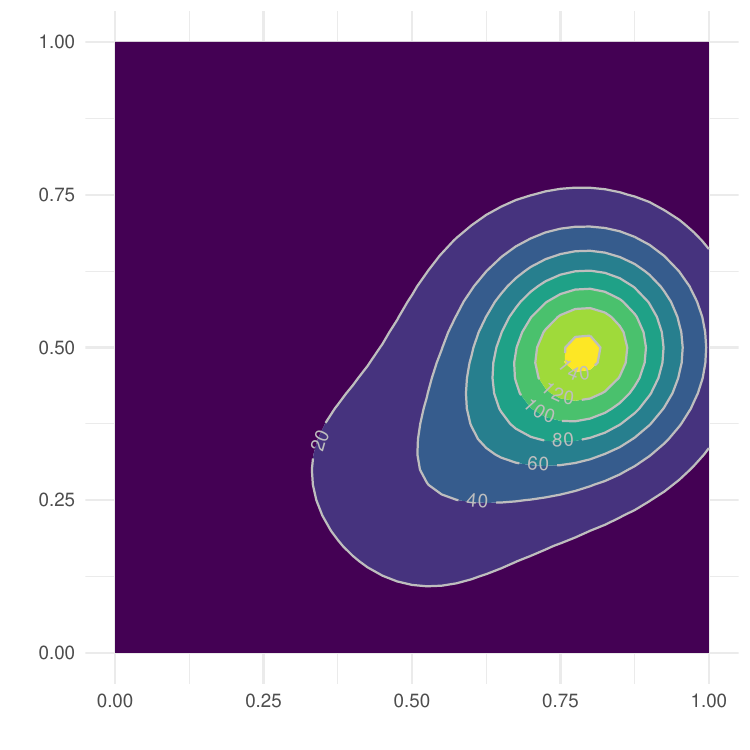}
    \end{minipage}
    \begin{minipage}{0.32\textwidth}
        \centering
        \includegraphics[width=\linewidth]{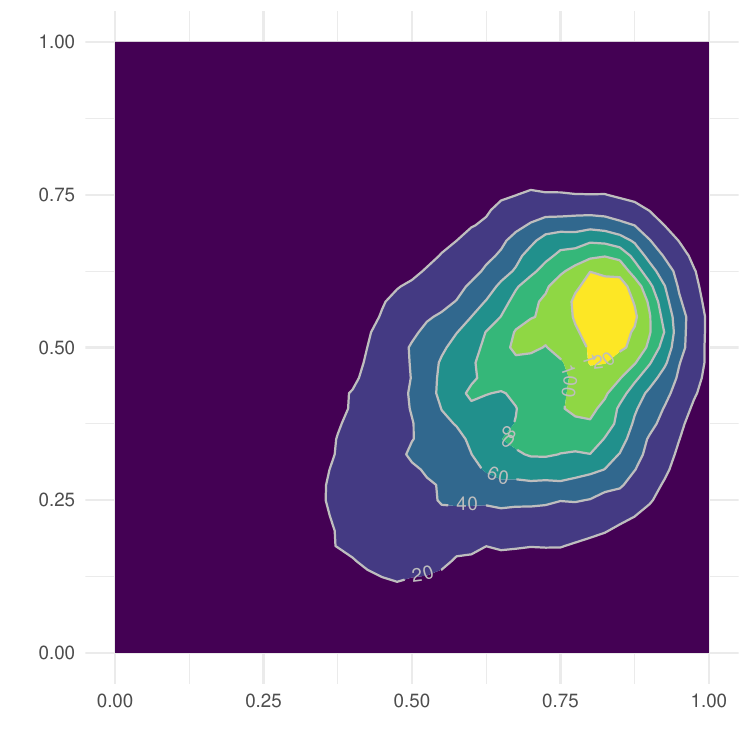}
    \end{minipage}
    \begin{minipage}{0.32\textwidth}
        \centering
        \includegraphics[width=\linewidth]{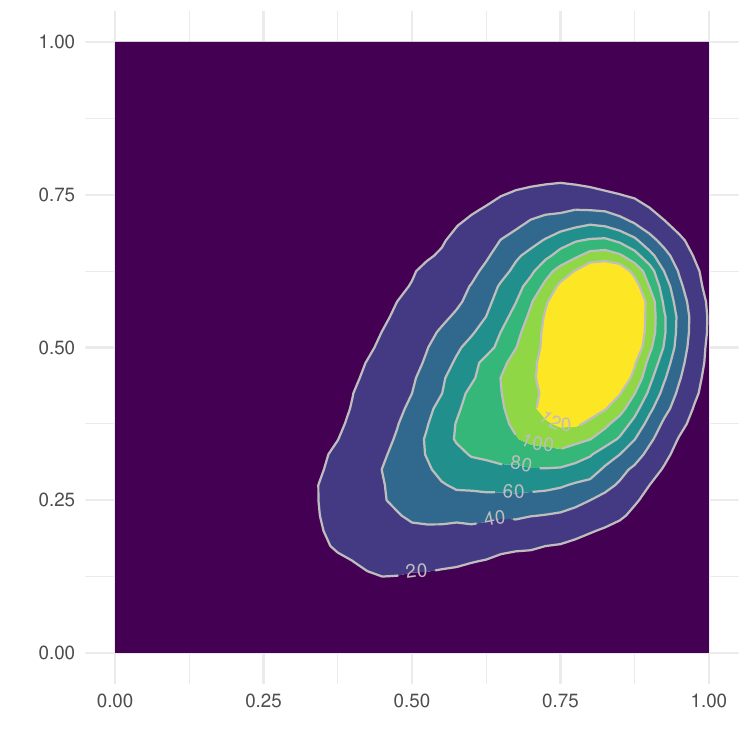}
    \end{minipage}
    \caption*{Time = 0.575}
\end{subfigure}

\vspace{-0.5em}

\begin{subfigure}{1\textwidth}
    \captionsetup{skip=1pt}
    \begin{minipage}{0.32\textwidth}
        \centering
        \includegraphics[width=\linewidth]{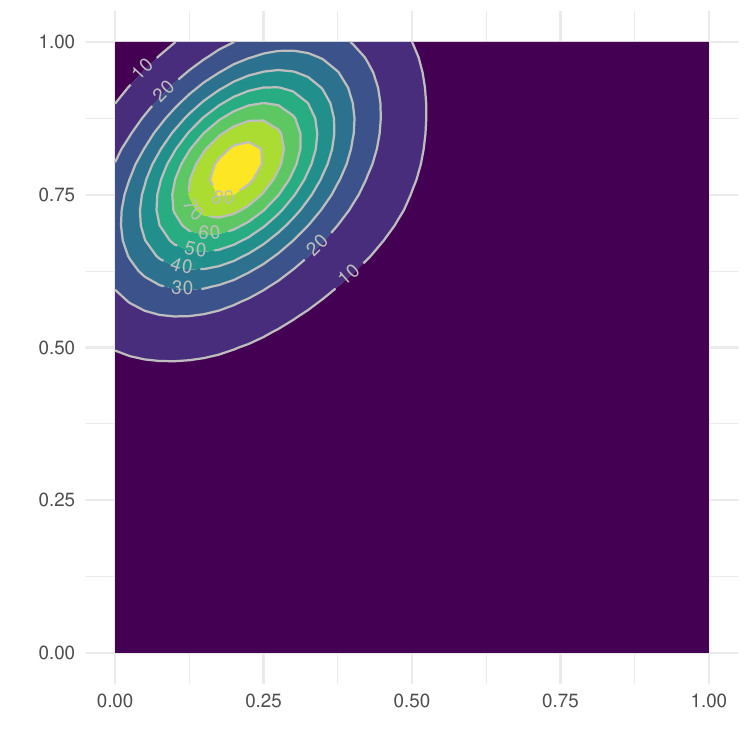}
    \end{minipage}
    \begin{minipage}{0.32\textwidth}
        \centering
        \includegraphics[width=\linewidth]{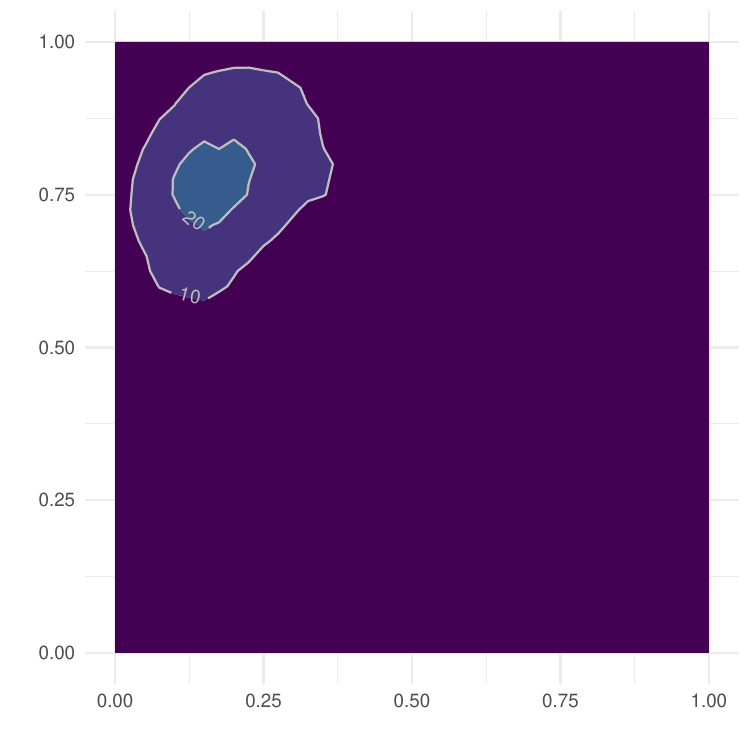}
    \end{minipage}
    \begin{minipage}{0.32\textwidth}
        \centering
        \includegraphics[width=\linewidth]{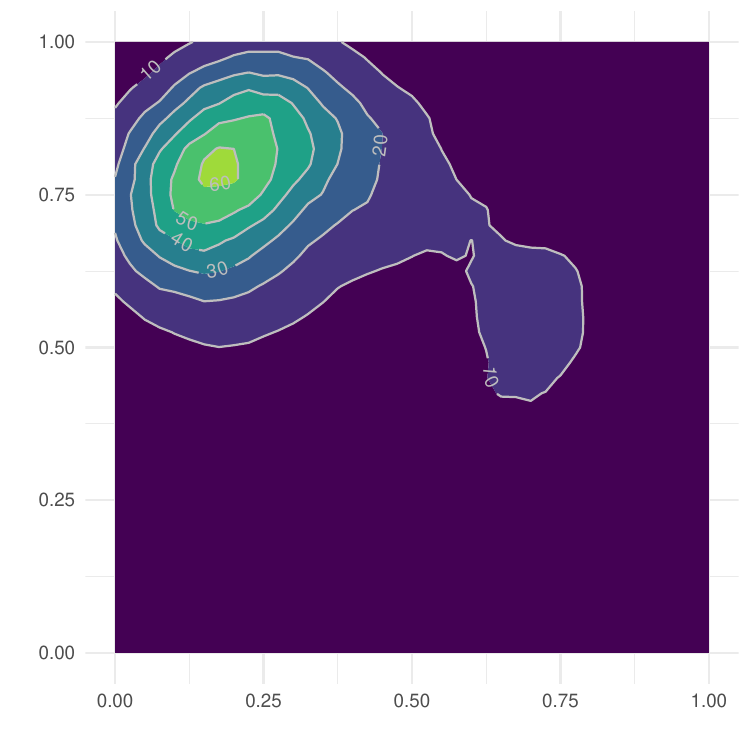}
    \end{minipage}
    \caption*{Time = 0.875}
\end{subfigure}

\caption{\review{Intensity estimation for the simulated spatio-temporal point pattern. Columns correspond to the true intensity function (left), the posterior mean obtained using the separable model (Model C, center), and the posterior mean from the non-separable model (Model D, right). Rows represent three key time points in the process: 0.225 (top), 0.575 (middle), and 0.875 (bottom).}}
\label{fig:sim_inten}
\end{figure}

\begin{figure}[htp]

\vspace{-0.5em}

\captionsetup{skip=5pt}
\centering
\textbf{Absolute value of temporal derivative of the intensity function}\par\medskip

\vspace{-0.3em}

\begin{minipage}{0.32\textwidth}
    \centering
    \textbf{True Values}
\end{minipage}
\begin{minipage}{0.32\textwidth}
    \centering
    \textbf{Model C}
\end{minipage}
\begin{minipage}{0.32\textwidth}
    \centering
    \textbf{Model D}
\end{minipage}

\vspace{0.3em}

\begin{subfigure}{1\textwidth}
    \captionsetup{skip=1pt}
    \begin{minipage}{0.32\textwidth}
        \centering
        \includegraphics[width=\linewidth]{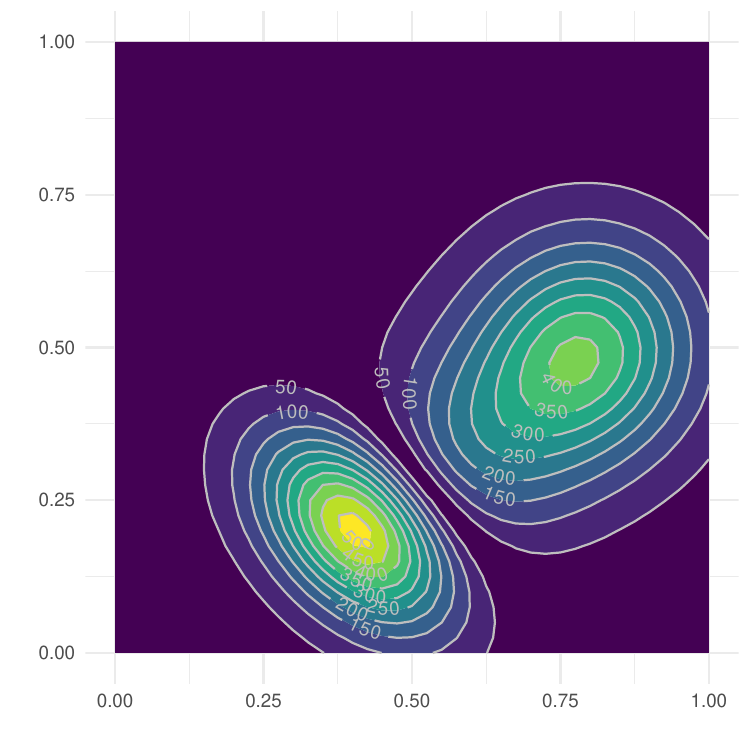}
    \end{minipage}
    \begin{minipage}{0.32\textwidth}
        \centering
        \includegraphics[width=\linewidth]{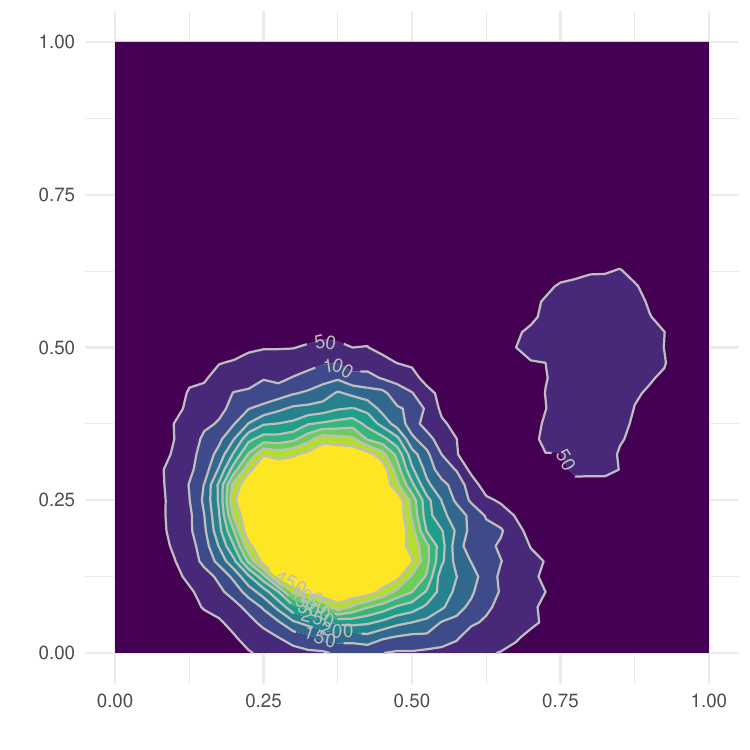}
    \end{minipage}
    \begin{minipage}{0.32\textwidth}
        \centering
        \includegraphics[width=\linewidth]{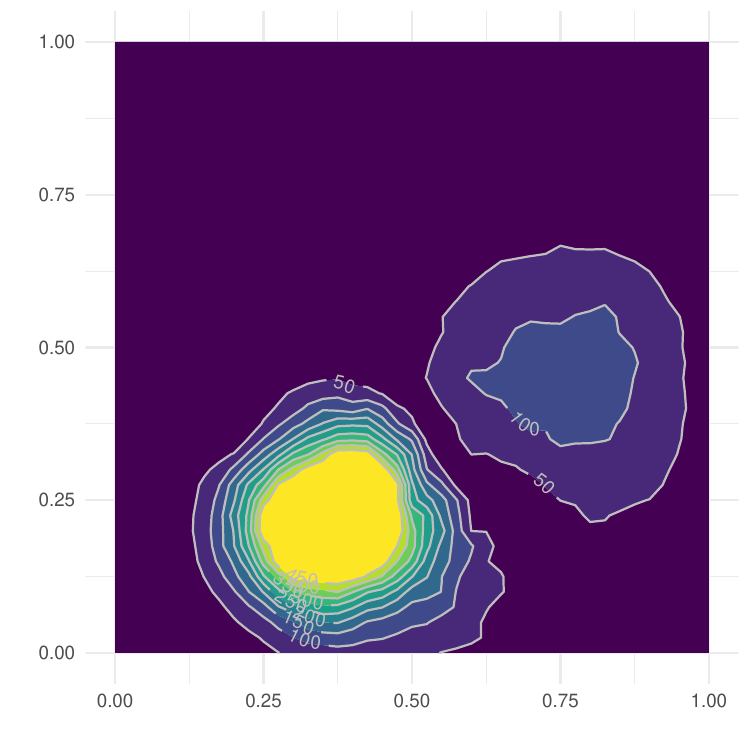}
    \end{minipage}
    \caption*{Time = 0.225}
\end{subfigure}

\vspace{-0.5em}

\begin{subfigure}{1\textwidth}
    \captionsetup{skip=1pt}
    \begin{minipage}{0.32\textwidth}
        \centering
        \includegraphics[width=\linewidth]{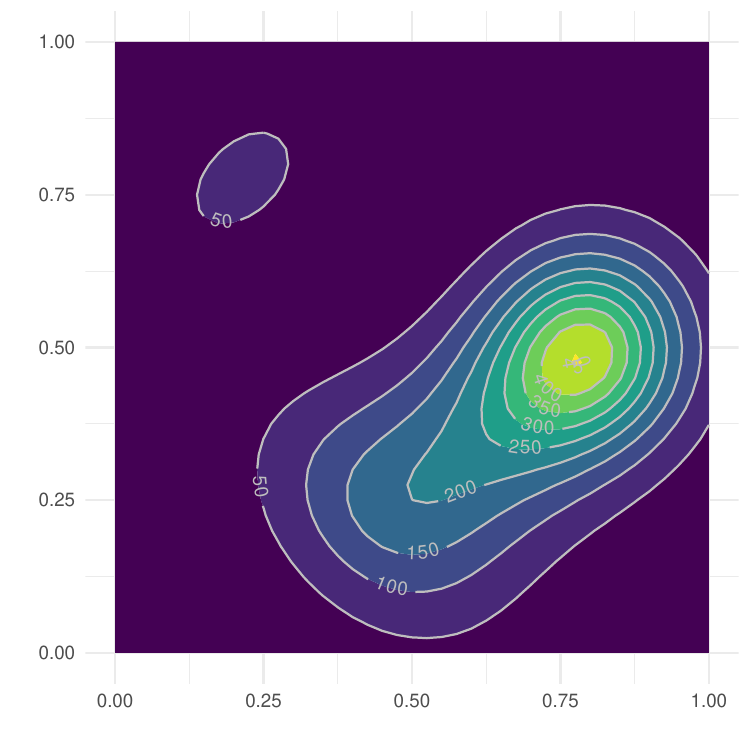}
    \end{minipage}
    \begin{minipage}{0.32\textwidth}
        \centering
        \includegraphics[width=\linewidth]{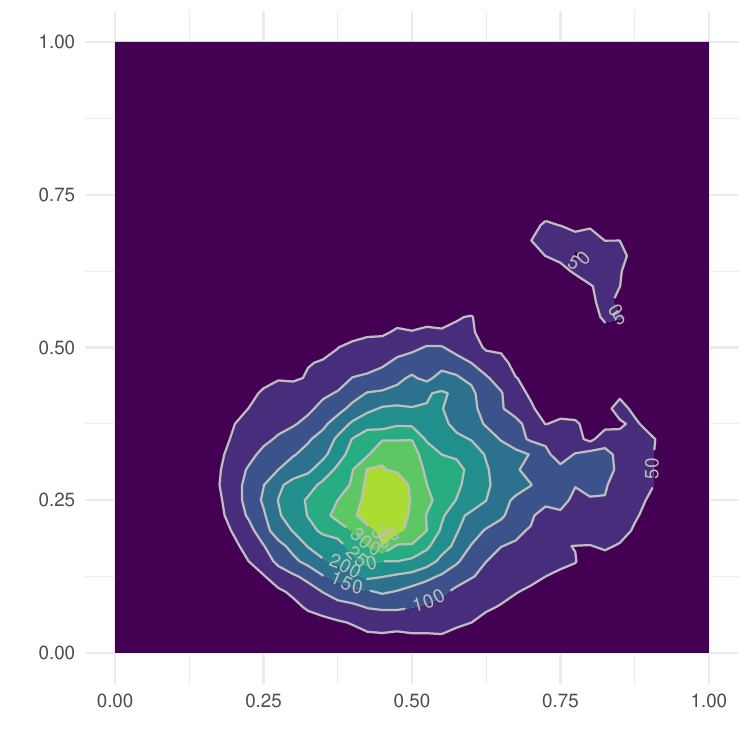}
    \end{minipage}
    \begin{minipage}{0.32\textwidth}
        \centering
        \includegraphics[width=\linewidth]{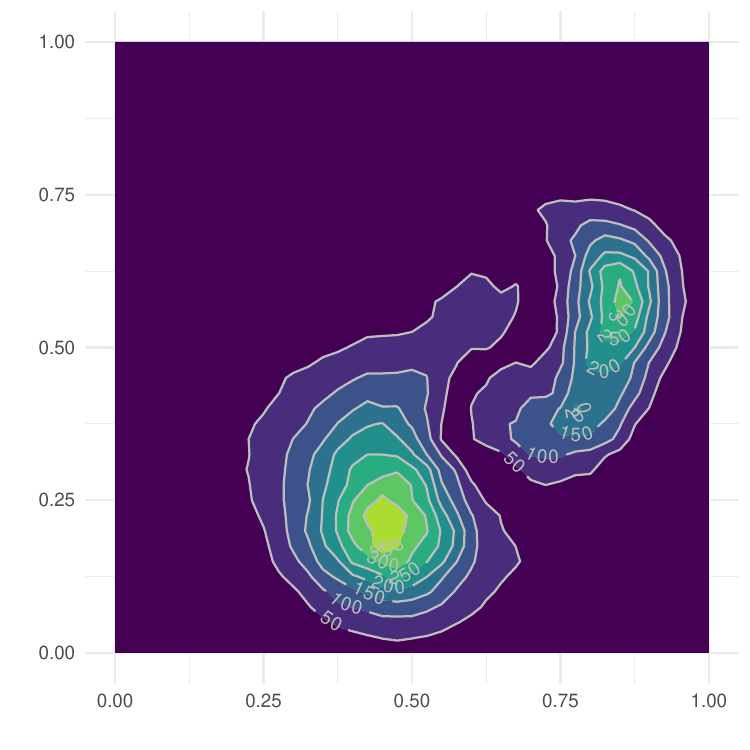}
    \end{minipage}
    \caption*{Time = 0.575}
\end{subfigure}

\vspace{-0.5em}

\begin{subfigure}{1\textwidth}
    \captionsetup{skip=1pt}
    \begin{minipage}{0.32\textwidth}
        \centering
        \includegraphics[width=\linewidth]{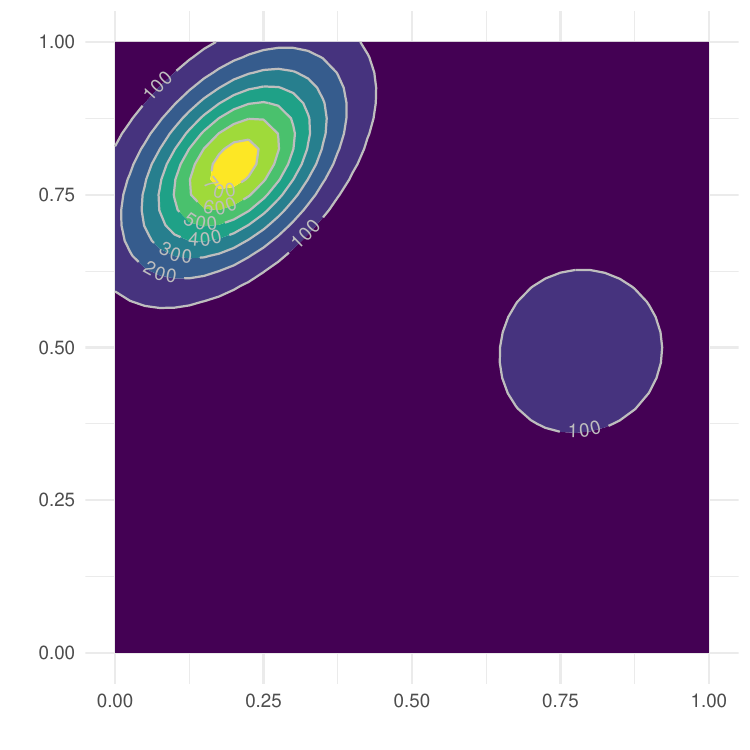}
    \end{minipage}
    \begin{minipage}{0.32\textwidth}
        \centering
        \includegraphics[width=\linewidth]{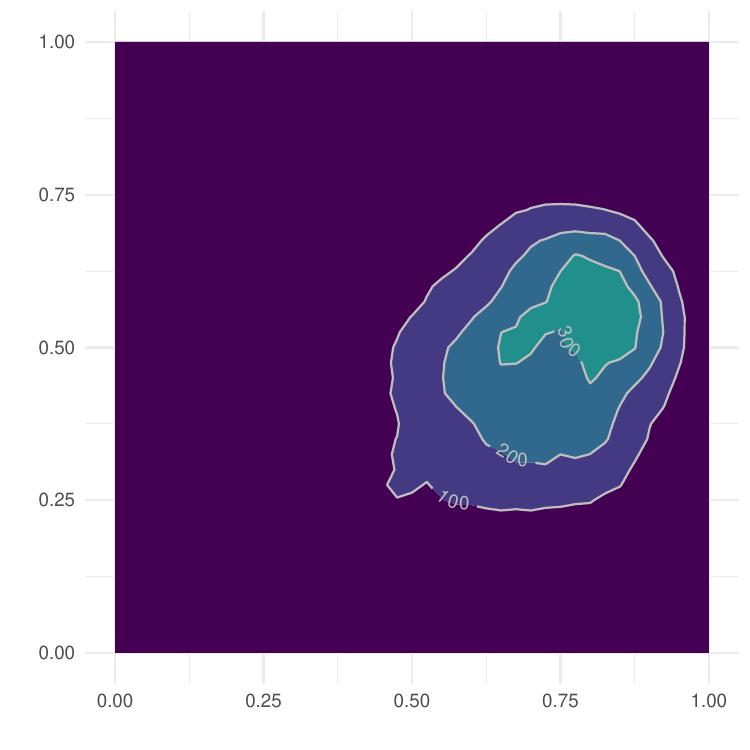}
    \end{minipage}
    \begin{minipage}{0.32\textwidth}
        \centering
        \includegraphics[width=\linewidth]{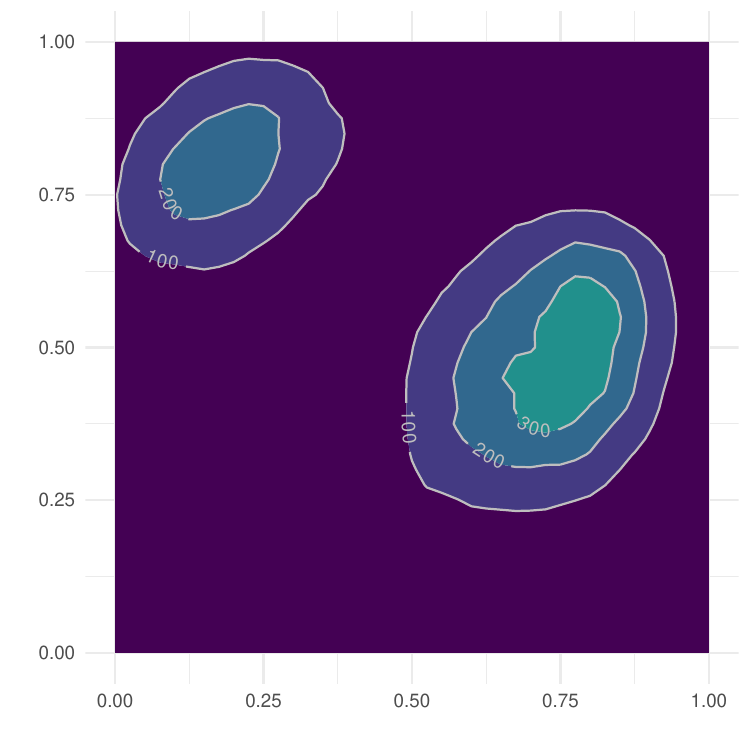}
    \end{minipage}
    \caption*{Time = 0.875}
\end{subfigure}

\caption{\review{Absolute value of the temporal derivative of the intensity function. Columns correspond to the true derivative (left), the approximation obtained using the separable model (Model C, center), and the approximation from the non-separable model (Model D, right). Rows represent three key time points in the process: 0.225 (top), 0.575 (middle), and 0.875 (bottom).}}
\label{fig:sim_temp}
\end{figure}

\begin{figure}[htp]

\vspace{-0.5em}

\captionsetup{skip=5pt}
\centering
\textbf{Norm of the spatial gradient of the intensity function}\par\medskip

\vspace{-0.3em}

\begin{minipage}{0.32\textwidth}
    \centering
    \textbf{True Values}
\end{minipage}
\begin{minipage}{0.32\textwidth}
    \centering
    \textbf{Model C}
\end{minipage}
\begin{minipage}{0.32\textwidth}
    \centering
    \textbf{Model D}
\end{minipage}

\vspace{0.3em}

\begin{subfigure}{1\textwidth}
    \captionsetup{skip=1pt}
    \begin{minipage}{0.32\textwidth}
        \centering
        \includegraphics[width=\linewidth]{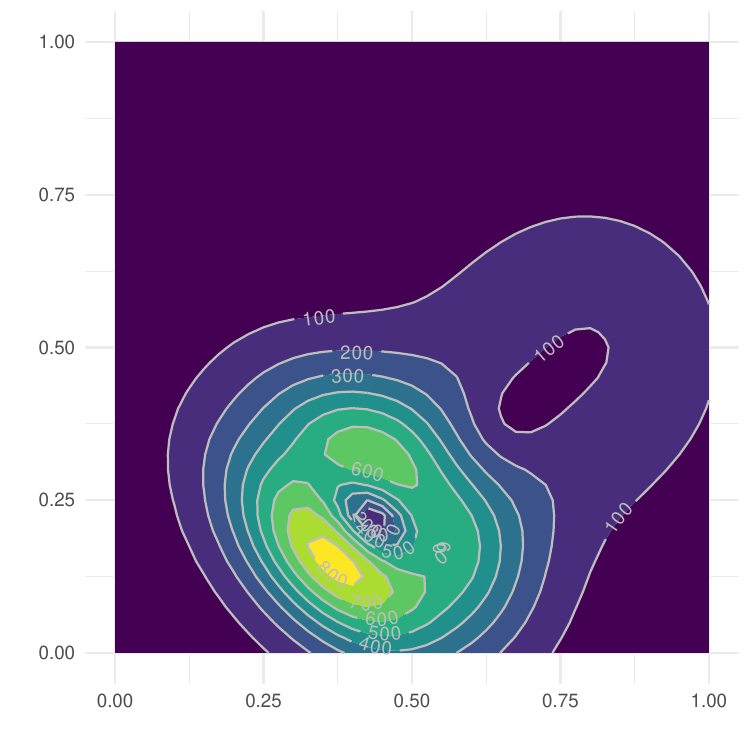}
    \end{minipage}
    \begin{minipage}{0.32\textwidth}
        \centering
        \includegraphics[width=\linewidth]{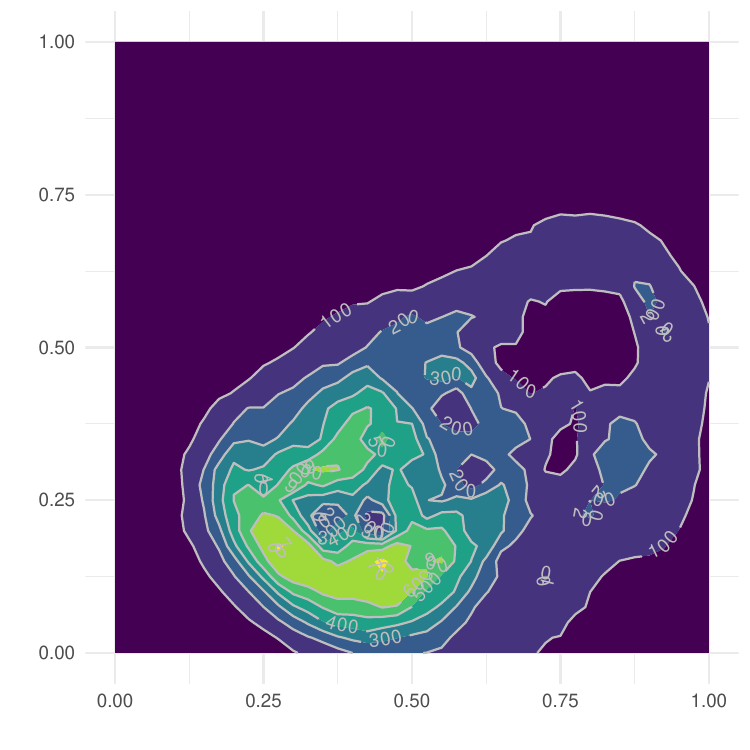}
    \end{minipage}
    \begin{minipage}{0.32\textwidth}
        \centering
        \includegraphics[width=\linewidth]{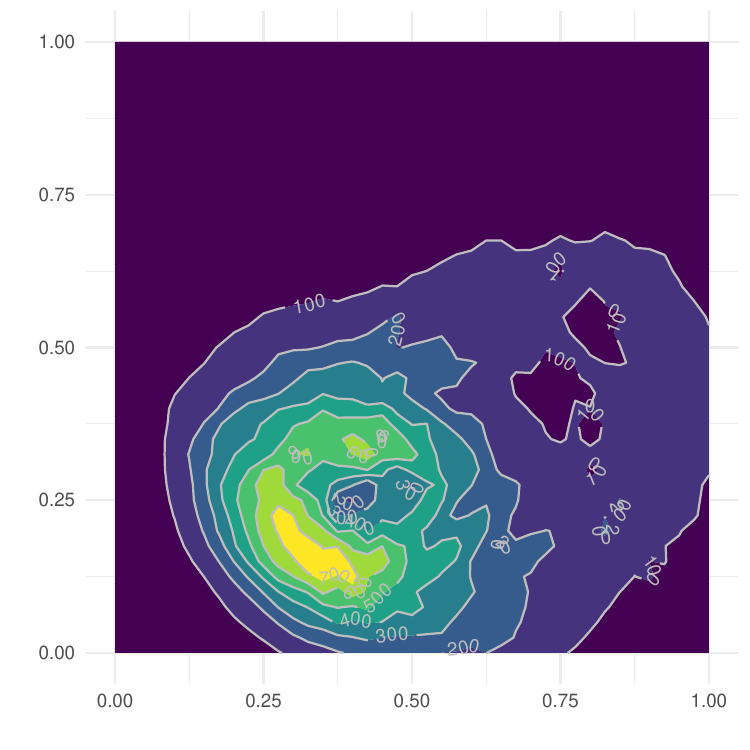}
    \end{minipage}
    \caption*{Time = 0.225}
\end{subfigure}

\vspace{-0.5em}

\begin{subfigure}{1\textwidth}
    \captionsetup{skip=1pt}
    \begin{minipage}{0.32\textwidth}
        \centering
        \includegraphics[width=\linewidth]{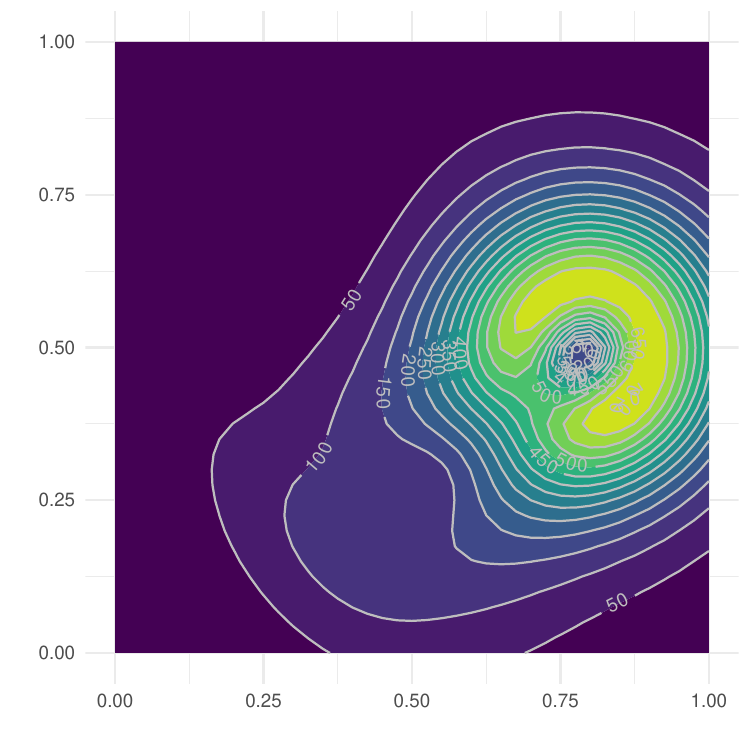}
    \end{minipage}
    \begin{minipage}{0.32\textwidth}
        \centering
        \includegraphics[width=\linewidth]{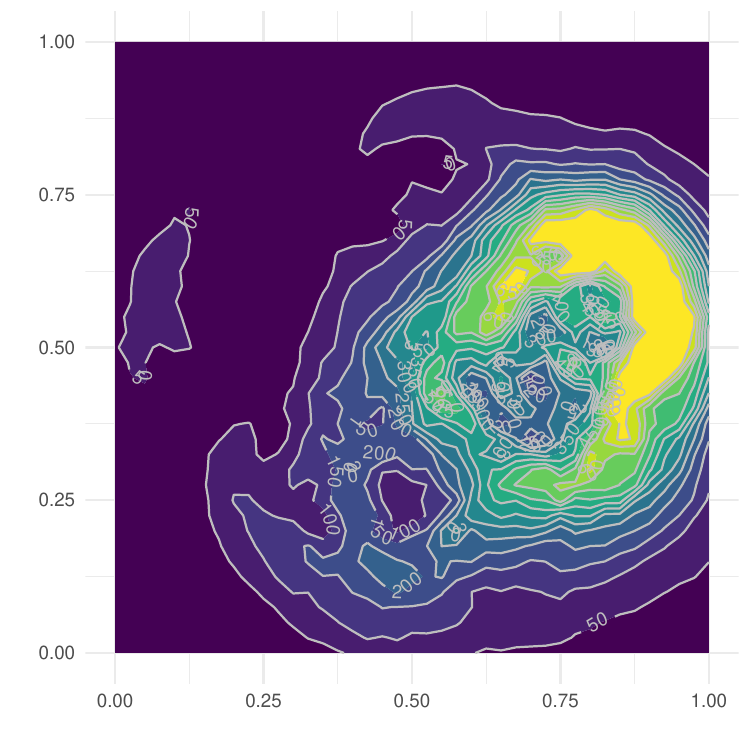}
    \end{minipage}
    \begin{minipage}{0.32\textwidth}
        \centering
        \includegraphics[width=\linewidth]{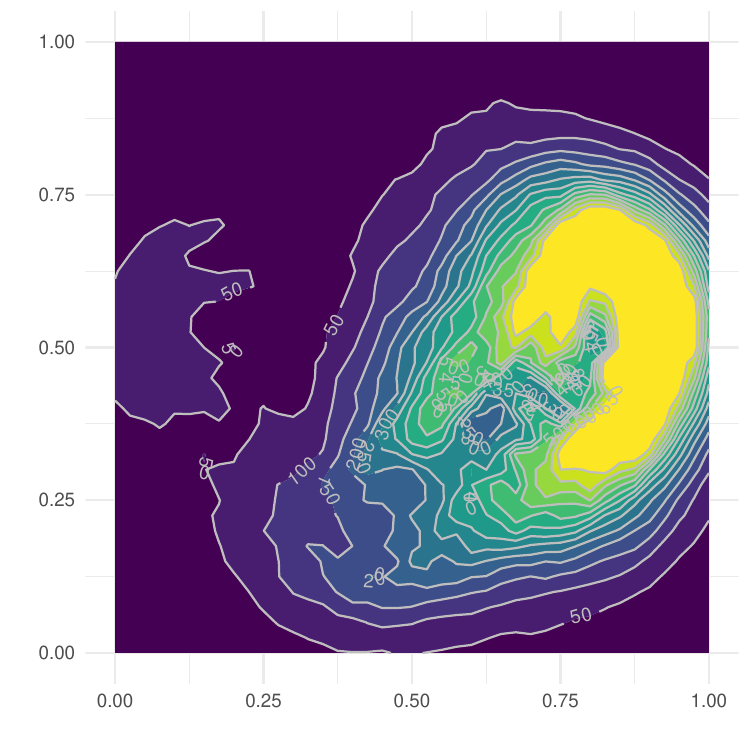}
    \end{minipage}
    \caption*{Time = 0.575}
\end{subfigure}

\vspace{-0.5em}

\begin{subfigure}{1\textwidth}
    \captionsetup{skip=1pt}
    \begin{minipage}{0.32\textwidth}
        \centering
        \includegraphics[width=\linewidth]{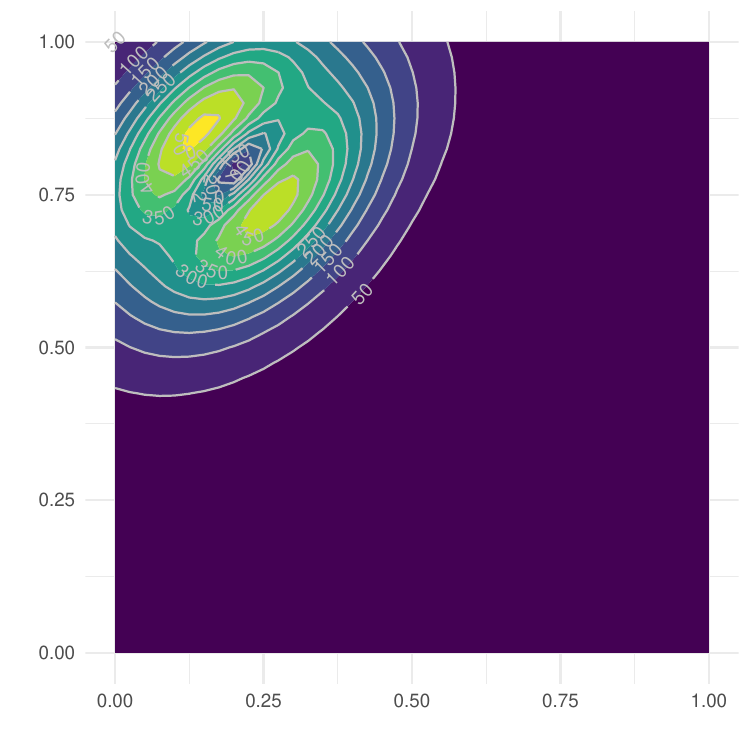}
    \end{minipage}
    \begin{minipage}{0.32\textwidth}
        \centering
        \includegraphics[width=\linewidth]{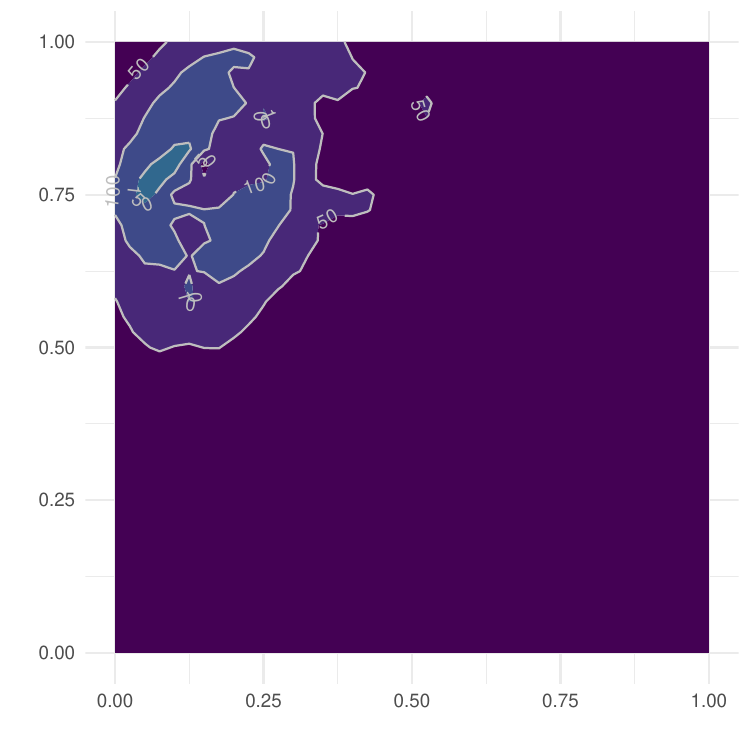}
    \end{minipage}
    \begin{minipage}{0.32\textwidth}
        \centering
        \includegraphics[width=\linewidth]{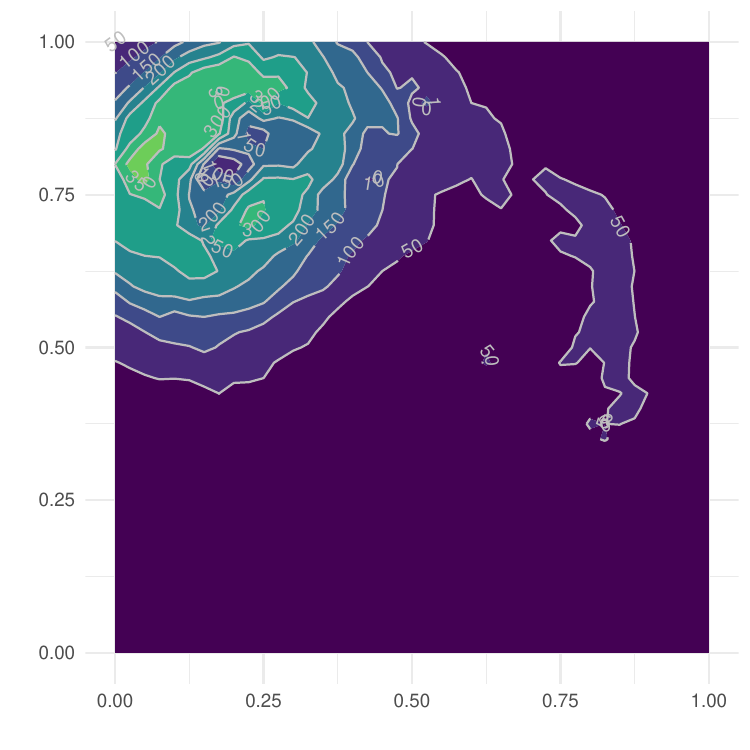}
    \end{minipage}
    \caption*{Time = 0.875}
\end{subfigure}

\caption{\review{Norm of the spatial gradient of the intensity function. Columns correspond to the true gradient norm (left), the approximation obtained using the separable model (Model C, center), and the approximation from the non-separable model (Model D, right). Rows represent three key time points in the process: 0.225 (top), 0.575 (middle), and 0.875 (bottom).}}
\label{fig:sim_norm}
\end{figure}

\review{This appendix presents the results of modeling the simulated spatio-temporal point pattern with the intensity function defined in (\ref{ecu:inten_sim}). Figures \ref{fig:sim_inten}, \ref{fig:sim_temp}, and \ref{fig:sim_norm} show, respectively, the intensity function, the absolute value of its temporal derivative, and the norm of its spatial gradient at time points $0.225$, $0.575$, and $0.875$, for $\lambda_0 = 5$. Each figure is organized by columns, with the first column showing the ground truth, and the second and third columns presenting the results from Models C and D, respectively.

Figure \ref{fig:sim_inten} compares the true intensity function with the posterior mean from each model. As expected, both models capture the temporal evolution of the intensity. However, Model D (non-separable model) in the third column generally provides a better approximation with smoother contour lines than Model C 
(separable model) in the second column. This smoothness is particularly important as it reduces noise in derivative computations, improving velocity approximations.  

Figure \ref{fig:sim_temp} presents the absolute values of the true and approximated temporal derivatives. While the approximations exhibit slightly more noise compared to the intensity estimates, both models successfully capture the overall trend. Notably, the non-separable model (Model D) provides a better fit than the separable model (Model C), particularly at times $0.575$ and $0.875$, reinforcing its advantage in derivative estimation.  

Finally, Figure \ref{fig:sim_norm} examines the spatial gradient norm for both the true and estimated values. This estimation exhibits even more noise than the temporal derivative approximation due to the additional step of combining spatial derivatives to compute the gradient norm. Despite this, both models capture the overall shape and mode of the true values. Once again, the non-separable model (Model D) demonstrates a better fit and a smoother surface compared to the separable model (Model C), further supporting its effectiveness in velocity estimation. }

\end{appendices}

\end{document}